\documentclass[prl,a4paper,twocolumn,superscriptaddress,longbibliography
]{revtex4-2}
\usepackage{multirow}
\usepackage{amssymb}
\usepackage{amsmath}
\usepackage{amsfonts}
\usepackage{graphicx}
\usepackage{bm}
\usepackage{multirow}
\usepackage{float}
\usepackage{relsize}
\usepackage{cmap}
\usepackage{bbold}
\usepackage{physics}
\usepackage{setspace}
\usepackage{cancel}
\usepackage{calligra}
\usepackage[dvipsnames]{xcolor}

\usepackage[colorlinks,citecolor=blue,linkcolor=red,urlcolor=blue]
{hyperref}

\definecolor{myGray}{gray}{0.9}

\usepackage[final]{pdfpages}
\usepackage{pgffor}

\makeatletter
\AtBeginDocument{\let\LS@rot\@undefined}
\makeatother

\DeclareMathOperator{\diag}{diag}

\begin{document}

\title{Dissipation-induced nonlocal phase-coherent transport in a topological superconductor}
	
	\author{S.\,V.\, Aksenov}
    \email{e-mail: asv@itp.ac.ru}
    \affiliation{
		L. D. Landau Institute for Theoretical Physics, 142432 Chernogolovka, Russia}
    \affiliation{Kirensky Institute of Physics, Federal Research Center KSC SB RAS, 660036 Krasnoyarsk, Russia}
    \affiliation{National Research Nuclear University MEPhI (Moscow Engineering
		Physics Institute), 115409 Moscow, Russia}
	\author{M.\,S.\, Shustin}
\affiliation{
		L. D. Landau Institute for Theoretical Physics, 142432 Chernogolovka, Russia}
    \affiliation{Kirensky Institute of Physics, Federal Research Center KSC SB RAS, 660036 Krasnoyarsk, Russia}
    \affiliation{National Research Nuclear University MEPhI (Moscow Engineering
		Physics Institute), 115409 Moscow, Russia}
    \author{I.\,S.\, Burmistrov}
    \affiliation{
		L. D. Landau Institute for Theoretical Physics, 142432 Chernogolovka, Russia}
	\affiliation{Laboratory for Condensed Matter Physics, HSE University, 101000 Moscow, Russia}
	
	\date{\today }
	
	\begin{abstract}
    Nonlocal phase-coherent transport via Majorana bound states, so-called 
teleportation, requires Coulomb blockade to suppress local Andreev reflection. 
We show that dissipation can replace the need for Coulomb blockade. We 
study a Kitaev chain coupled to a Markovian bath and identify a regime where 
the magnitude of nonlocal conductance contribution reach the local one. In 
this regime, the lowest-energy subspace is  maximally mixed, reminiscent 
of a directly-grounded nontrivial topological superconductor. Phase coherence is confirmed by 
Aharonov-Bohm oscillations. Our results establish dissipation as a tool for 
controlling quantum teleportation in Majorana systems.
	\end{abstract}	
	
	\maketitle

The search for Majorana bound states (MBS) in various solid-state platforms \cite{moore-91, volovik-99, fu-08, lutchyn-10, oreg-10, nadj-perge-13, backens-22, yazdani-23} remains an intriguing long-standing 
problem and their unambiguous detection is still an open challenge (see, e.g.,
\cite{dassarma-23, frolov-23, legg-26, microsoft-26}). Several years ago, these studies gained renewed momentum with the implementation of 
the minimal Kitaev chain model~\cite{kitaev-01, leijnse-12, sau-12} in 
superconducting quantum dot systems~\cite{dvir-23, tenhaaf-24}. The 
measurement of ground-state parity in these platforms has recently been 
demonstrated using quantum capacitance \cite{vanloo-26,zhang-26b}. Soon after, 
coherent coupling between two such systems, forming a Majorana parity qubit 
\cite{leijnse-12,tsintzis-24,pino-24}, was achieved \cite{zatelli-26}. 

An important concept that emerged in this context is that of ``quantum 
teleportation" \cite{fu-10, nitsch-25}. Specifically, in the Coulomb blockade 
regime, resonant local Andreev reflection (LAR) on a Majorana mode (MM)
\cite{bolech-07,law-09,flensberg-10} becomes energetically unfavorable, while 
phase-coherent transport between two Fermi reservoirs dominates, analogous to 
the normal-lead/quantum-dot/normal-lead case. This process enables readout of 
the topological qubit state via the transmission phase shift in electron 
teleportation between well-separated MMs \cite{vijay-16,karzig-17,plugge-17}. 

At the same time, a pressing problem in the field remains quasiparticle 
poisoning, i.e., the impact of dissipation on the topological qubit 
\cite{goldstein-11, budich-12, rainis-12,albrecht-17,karzig-21,vanloo-26, zhang-26b,zatelli-26}. 
The poisoning may originate from external fermionic reservoirs. The standard 
strategy to mitigate these effects is to introduce a strong charging energy 
\cite{vijay-16,karzig-17,plugge-17}. However, the effects of poisoning on the 
Coulomb-blockaded Majorana qubit have not been thoroughly studied yet
\cite{bhattacharyya-26}.

On the other hand, it is known that dissipation can actively stabilize 
topological phases and maintain quantum coherence \cite{diehl-11, krauter-11, 
bardyn-12, otterbach-14, gong-17, goldstein-19, gogoi-26}. This raises an intriguing 
question: could it, contrary to conventional wisdom \cite{stern-90,datta-95}, be harnessed to enable coherent nonlocal transport through MBS and provide 
control over their fermion parity? More specifically, can the nonlocal nature 
of MBS be exploited in a topological superconductor interacting with an 
engineered nonlocal bath to mediate transport features akin to quantum 
teleportation?

\begin{figure}[b]
\centerline{\includegraphics[width=0.90\columnwidth]{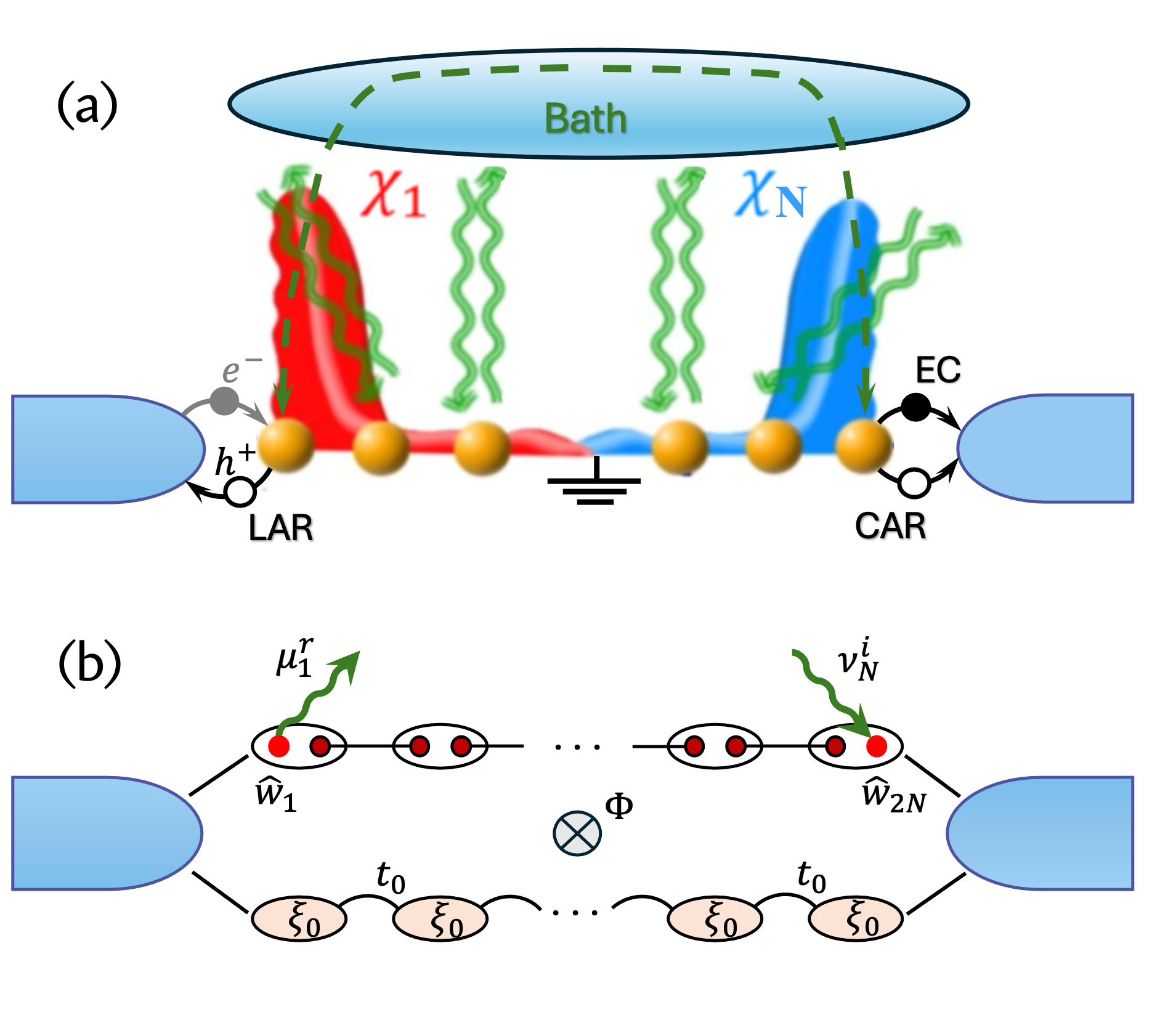}}
\caption{(a) Sketch of a dissipative directly-grounded Kitaev chain between two normal leads. The Majorana wave functions $\chi_{1,N}$ are depicted by red and blue colors. The bath influence is schematically shown by green wavy and dashed lines. The LAR, CAR, and EC processes are sketched with initial process of electron tunneling shown in gray and the final processes shown in black lines with filled (for an electron) and empty (for a hole) circles. (b) Scheme of a setup for AB-type experiment. (see text)}
\label{model} 
\end{figure}

In this Letter we answer this question affirmatively and suggest a way to 
observe the nonlocal phase-coherent transport even in a directly grounded 
topological superconductor. To achieve this effect one has to employ a 
Markovian-type bath affecting the topological superconductor 
(Fig.~\ref{model}a). We demonstrate that, generically, the dissipation 
induced by the bath creates low-energy current-carrying states dwelling at 
both device edges. This is in stark contrast to the conventional 
dissipation-free case, in which the wave functions of MMs are localized 
at opposite ends of the topological superconducting wire. As a result, the 
elastic cotunneling (EC) and crossed Andreev reflection (CAR) contributions 
to the conductance become nonzero and distance independent. To prove the phase coherence of the nonlocal transport we compute the 
Aharonov-Bohm (AB) oscillations of the conductance in a two-arm setup 
(Fig.~\ref{model}b).
Also we establish the  link between the MBS parity and different transport regimes.

\noindent\textsf{\color{blue}Model.} We model the time dynamics of the system 
sketched in Fig.~\ref{model} by the Gorini-Kossakowski-Sudarshan-Lindblad 
(GKSL) equation for the density matrix $\hat{\rho}$,
\begin{equation}\label{Lindblad_eq}
d \hat{\rho}/d t = -i\,[\,\hat{H},\,\hat{\rho}\,] + \left(\,\hat{L}\hat{\rho} \hat{L}^\dagger - \{\hat{L}^\dagger\hat{L},\hat{\rho}\}/2\right).
\end{equation}
The unitary part of the evolution is governed by the Hamiltonian 
$\hat{H}=\hat{H}_s+\hat{H}_L+\hat{H}_T$. 
The first term describes the Kitaev chain Hamiltonian, which belongs to the 
BDI symmetry class,
\begin{equation}\label{HK}
\hat{H}_s = -\sum_{n=1}^N\mu\,\hat{c}_{n}^\dagger\hat{c}_n +\sum_{n=1}^{N-1}\left [ (t\hat{c}_{n}^\dagger + \Delta\hat{c}_n) \hat{c}_{n+1} + {\rm h.c.}\right] .
\end{equation}
Here $\hat{c}_{n}^\dagger$ ($\hat{c}_n$) is the creation (annihilation) operator 
at site $n$. The real parameters $\mu$, $t$, and $\Delta$ are the chemical 
potential, hopping amplitude, and pairing amplitude, respectively.
The noninteracting single-band leads are described by the Hamiltonian 
$\hat{H}_L=\sum_{ik}\xi_{i,k}\hat{d}_{i,k}^{\dagger}\hat{d}_{i,k}$, 
where $\hat{d}_{i,k}$ ($\hat{d}_{i,k}^{\dagger}$) annihilates (creates) an 
electron in the $i$-th lead with energy $\xi_{i,k}$ and wave vector $k$.
 In our analysis, we adopt the standard wide-band approximation assuming the constant density of states in the leads, $\rho_i=\sum_k \delta(\xi_{i,k})$.
The 
tunneling Hamiltonian 
$\hat{H}_T=\sum_{i=1,N}\sum_{k} t_{i} \hat{d}_{i,k}^{\dagger} \hat{c}_{i}+{\rm h.c.}$ 
describes tunneling between the $i$-th lead and the edge sites of the 
superconducting chain with amplitudes $t_i$. For convenience, each lead shares 
the index of the chain site ($i=1$ or $i=N$) to which it is attached. We assume that the tunneling rates $\Gamma_i=2\pi |t_i|^2 \rho_i \ll |t|, |\Delta|$.

In this work, we focus on jump operators $\hat{L}$ that are linear in fermionic 
creation and annihilation operators:
$\hat{L}=\sum_{n}\left(\mu_{n}\hat{c}_{n}+\nu_{n}\hat{c}_{n}^{\dagger}\right)$. This form of $\hat{L}$ corresponds to tunnel coupling of the superconducting system with a reservoir of free fermions \cite{dabbruzzo-21a,shustin-26}.
The amplitudes of electron loss $\mu_n=\mu_n^r+i\mu_n^i$ and gain 
$\nu_n=\nu_n^r+i\nu_n^i$ due to the interaction with the bath are assumed to be 
sufficiently weak: $|\mu_n|, |\nu_n| \ll |t|, |\Delta|$.

\noindent\textsf{\color{blue}Topological phase.} We are interested in the regime $t>0$, $\Delta<0$, and $|\mu|<2t$ for which the ground state of the Hamiltonian $H_s$ is in the topological phase characterized by a winding number 
$N_{\rm BDI} = \oint d\beta_k/(2\pi)\equiv -1$ where $\tan \beta_k = 2\Delta \sin k/(2t\cos k-\mu)$ \cite{Budich2013, Sato2017, valkov-22}. In this case, the Hamiltonian $H_s$ possesses a pair of MMs (constituting the MBS) localized near the opposite ends of the chain and characterized by the wave functions $\underline{\chi}_{1}, \,\underline{\chi}_{N} \in \mathbb{R}^{N}$ (see Fig.~\ref{model}). In the special case of a symmetric point of the Kitaev model ($\mu = 0$, $|t| = |\Delta|$) that is also known as a {\it sweet spot} and actively studied nowadays in systems of superconducting quantum dots \cite{leijnse-12,dvir-23,tenhaaf-24}, MMs are localized on the first and last sites of the chain, $\chi_{j,n}=\delta_{n,j}$ (see End Matter). 
\begin{figure*}[t]
\centerline{
    \includegraphics[width=0.3275\textwidth]{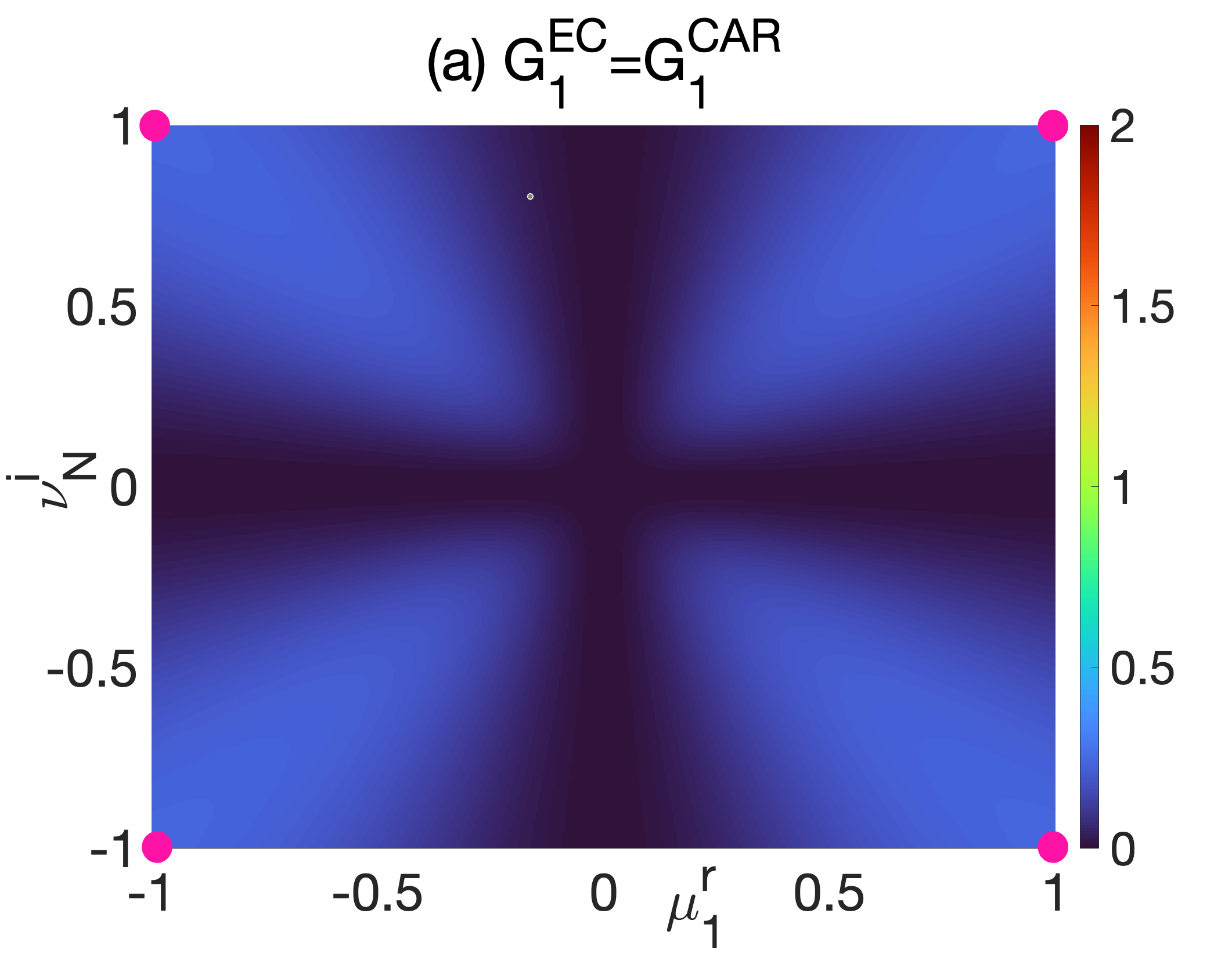}\quad
    \includegraphics[width=0.3275\textwidth]{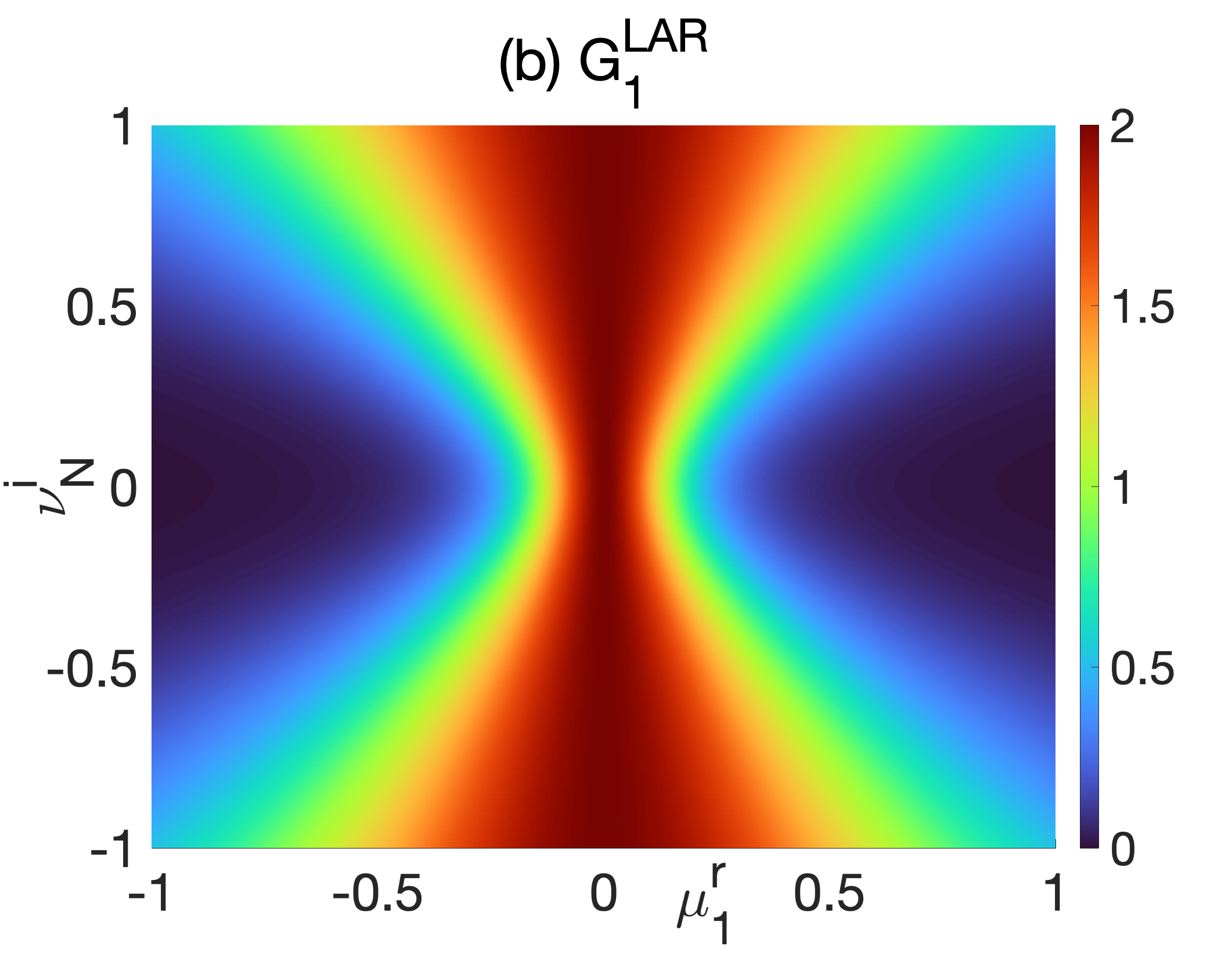}\quad
    \includegraphics[width=0.3275\textwidth]{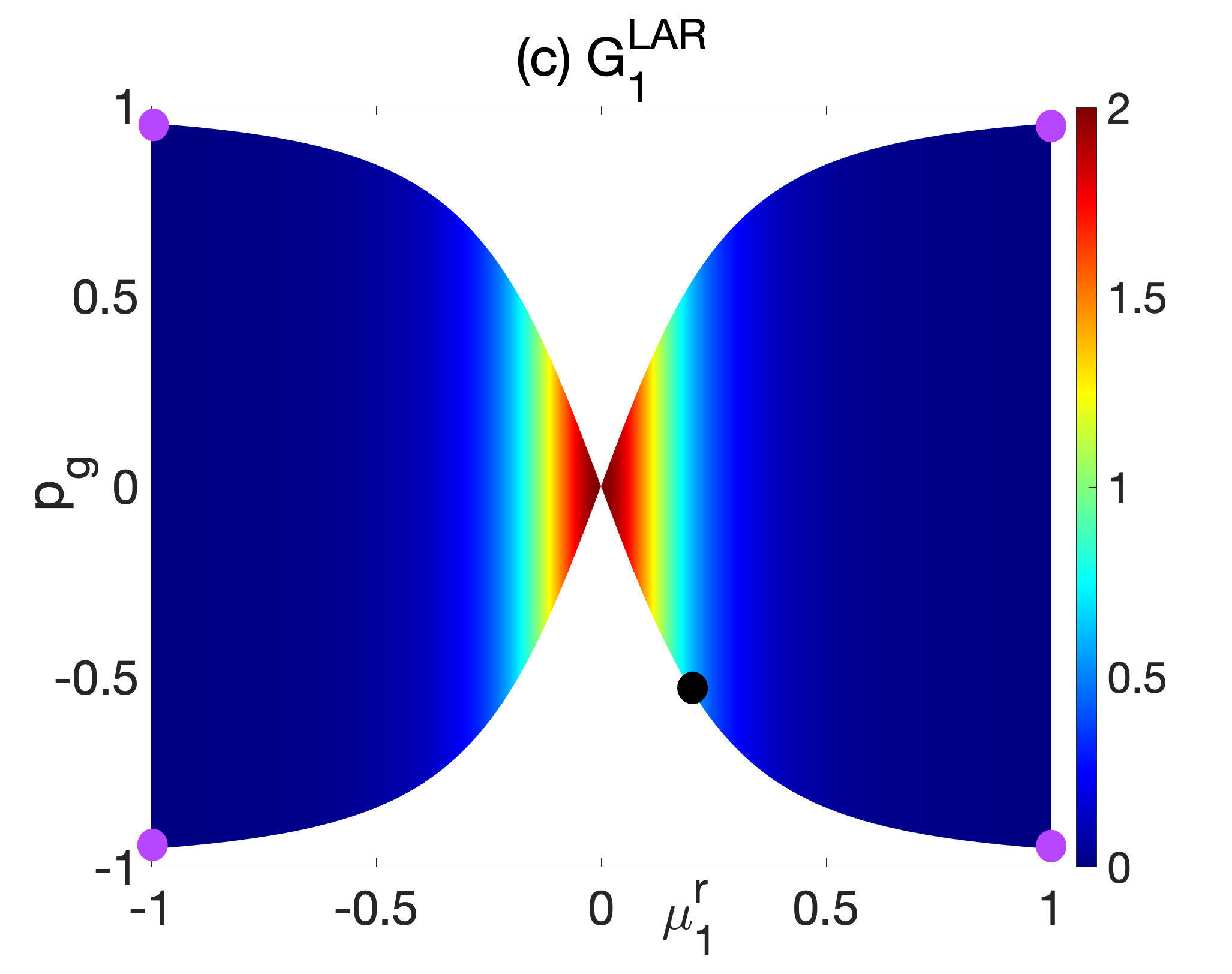}\quad
    }
\caption{The effect of dissipation on the left-lead conductance at zero bias voltage and at the sweet spot of the Kitaev chain model. ($\mu_{1}^{r}$, $\nu_{N}^{i}$)-maps of the nonlocal (panel a) and the local (panel b) contributions for the identically zero parity of the open MBS, $p_g=0$. Panel c:  ($\mu_{1}^{r}$, $p_g$)-map of the LAR conductance in the case of the identically zero nonlocal contributions, $G_1^{EC/CAR}=0$. Pink (violet) dots in panel a (c) indicate the regime of maximal nonlocal transport and $p_g=0$ (dissipative blockade and $p_g\approx\pm1$). Black dot in panel c corresponds to parameters used in Fig. \ref{G1_AB}b. 
Parameters used: $g_{1,N}=0.05$, $\nu_{1}=0$, $\mu_{N}=0$, $\mu_{1}^{i}=0$, and $\nu_{N}^{r/i}=0$.
}
\label{G1_pt}
\end{figure*}

\noindent\textsf{\color{blue} Nonlocal electric transport.} We assume that each lead is biased by a voltage $V_j$, with $j=1,N$. A current 
$I_j$ then flows into the $j$-th lead. To compute it, we employ the extension 
of the Meir-Wingreen formula to dissipative systems, recently derived in Ref.~\cite{aksenov-26}. At zero temperature, the local conductance of the $j$-th 
lead decomposes into contributions from LAR, EC, CAR, and electron tunneling into 
the bath (B) (see Fig.~\ref{model} and End Matter):
\begin{equation}\label{G_conts}
G_{j}=\partial I_{j}/\partial V_{j}=G_{j}^{LAR}+G_{j}^{EC}+G_{j}^{CAR}+G_{j}^{B}.
\end{equation}
Here and below, conductance is measured in units of $G_{0}=e^2/h$.
For low voltage, $|V_j| \ll |t|, |\Delta|$, the transport is mediated by the MBS. 
Then, treating the coupling of the system to the leads and the bath as a perturbation, the energy splitting and the structure of subgap excitations can be analyzed to leading order by projecting the Lindbladian onto the MBS subspace. As a result, we find
\begin{equation}
\begin{split}
G_{j}^{LAR} & = 2{g}_j^2\frac{{V}_j^2/4+\left({g}_{\bar{j}}+|\bm{b}_{\bar{j}}|^2\right)^2}{\left({V_j^2}/4+\beta_{1}^2\right)\left(V_j^2/4+\beta_{N}^2\right)},  \\
G_j^{EC} & = G_j^{CAR} =  \frac{{g}_1{g}_N [\bm{b}_{1}\times \bm{b}_{N}]^2}{\left({V_j^2}/4+\beta_{1}^2\right)\left(V_j^2/4+\beta_{N}^2\right)}, \\
G_j^B  
& = \frac{|\bm{b}_j|^2}{{g}_j} G_{j}^{LAR}- 2\left (2 +\frac{|\bm{b}_{\bar{j}}|^2}{{g}_{\bar{j}}}\right )G_{j}^{EC} .
\end{split}\label{G_terms}
\end{equation}
Here $\bar{j}=N,1$ if $j=1,N$ and $g_{j}=\Gamma_j \chi^2_{j,j}/4$ denotes the local tunneling rate that accounts for MMs probability density at the chain edges. Two-dimensional vectors  
\begin{equation}\label{sa_vecs}
{\bm{b}}_{j} =  \{(\,\underline{\mu}^r +s_j \underline{\nu}^r\,) \cdot \underline{\chi}_{j}, \quad (\,\underline{\mu}^i +s_j  \underline{\nu}^i\,) \cdot \underline{\chi}_{j} \}/2 ,
\end{equation}
where $s_j=\delta_{j,1}-\delta_{j,N}$, describe hybridization between the MBS wave functions and dissipative fields. The energies of the lowest lying states read
\begin{equation}\label{eq:beta:1N}
 \beta_{1,N} {=} \sum_j \frac{g_j{+}|\bm{b}_j|^2}{2} {\pm} \Bigl \{\Bigl [\sum_j \frac{s_j (g_j{+}|\bm{b}_j|^2)}{2}\Bigr ]^2{+} [\bm{b}_{1}{\times}\bm{b}_{N}]^2\Bigr \}^{\frac12} .    
\end{equation}
\color{black}We mention the coincidence of EC and CAR contributions to the conductance, cf. Eq.~\eqref{G_terms}. This fact is well-known in the absence of dissipation~\cite{nilsson-08,tikhonov-26}. 

In the absence of dissipation, only the $G_j^{LAR}$ contribution to the 
conductance survives, reaching its maximal value of $2$ at zero bias. 
Electric transport is thus fully local, as expected when MMs do not overlap \cite{bolech-07,law-09}. The presence of dissipation completely changes this 
picture. In addition to the positive bath contribution, $G_j^{B}>0$, dissipation 
unexpectedly facilitates nonlocal transport: 
the $G_j^{EC}$ and $G_j^{CAR}$ contributions becomes nonzero provided the vectors $\mathbf{b}_{1}$ and 
$\mathbf{b}_{N}$ are not parallel. 
In this case, the eigenstates of the 
Liouvillian in the presence of the bath become linear combinations of the 
Majorana wave functions $\underline{\chi}_{1,N}$ and, consequently, are 
localized at both ends of the chain. Accordingly, the realization of nonlocal transport requires that (i) the bath acts on both edges of the device, and (ii) the dissipative fields 
$\underline{\mu}$ and $\underline{\nu}$ are neither purely real nor purely 
imaginary.

The nonlocal conductance contributions are bounded from above by 
$G_{j}^{EC}=G_{j}^{CAR}\leqslant 1/4$. Perfect ``teleportation", with 
$G_{j}^{EC}+G_{j}^{CAR}=1$, is therefore impossible. The maxima of 
$G_{j}^{EC}$ and $G_{j}^{CAR}$ are attained at zero bias, $V_{j}=0$, for 
orthogonal vectors $\bm{b}_1\perp\bm{b}_N$, and in the symmetric setup 
$|b_{1,N}|^2 \equiv b^2$ and $g_{1,N}\equiv g$, with weak tunnel coupling to 
the leads, $g\ll b^2$: 
$G_{j}^{EC}=G_{j}^{CAR}\simeq (1-g/b^2)/4$. 
At the same time, $G_j^{LAR}$ is reduced to $(1+g/b^2)/2$, while the bath 
contribution is almost completely suppressed, $G_j^{B}\simeq g/(2b^2)$.

\noindent\textsf{\color{blue} Average fermion parity.} The conditions for the maximum of the nonlocal conductance contributions 
correspond to the situation when the eigenfunctions of the lowest-lying states 
have equal weight at the edges of the chain. 
It is useful 
to consider the reduced density matrix on the Fock subspace spanned by the 
low-lying states with energies $\beta_{1,N}$. Then, for the symmetric transport setup, $g_{1,N}\equiv g$ and 
at zero bias, we find (see End Matter)
\begin{equation}\label{rho}
\rho_{\rm red} {=} \frac{1}{2}
\begin{pmatrix}\,
    1 {+} p_g & 0 \\
    0 & 1 {-} p_g \,
\end{pmatrix}, \quad p_g {=}\frac{2\bm{b}_1\cdot \bm{b}_N}{2g{+}|\bm{b_1}|^2{+}|\bm{b_N}|^2} .
\end{equation}
The quantity $p_g$ is the average fermion parity in the two-dimensional 
subspace spanned by the lowest many-body states from the even- and odd-parity 
sectors (see End Matter). The excitations in this subspace have a finite 
lifetime due to the coupling to the leads and the bath. In other words, one 
can treat $p_g$ as the parity of the open MBS. 
In the absence of leads, $g=0$, the average fermion parity becomes 
$p = 2\bm{b}_1\cdot \bm{b}_N/(|\bm{b_1}|^2{+}|\bm{b_N}|^2)$. This reveals the meaning of the angle 
between $\bm{b}_1$ and $\bm{b}_N$: it controls the average fermion parity of 
the MBS hybridized by the bath in the absence of tunneling 
to the leads.

The  
density matrix \eqref{rho} is mixed in the presence of the 
leads. It is maximally mixed when $p_g{=}p{=}0$, i.e., when the eigenfunctions of 
the low-lying states are equally spread over both edges of the chain, thus 
facilitating maximal nonlocal transport. This sharply contrasts with the 
mechanism in Ref.~\cite{fu-10}, where nonlocal transfer was achieved by 
fixing the fermion parity. In this sense, our system supporting maximal 
nonlocal transport is conceptually closer to the ground state of a conventional 
isolated topological superconductor, for which the average fermion parity in 
the bulk is zero due to the fundamental degeneracy of the entanglement 
spectrum~\cite{vidal-03, turner-11}.

In the opposite limit, $p=\pm1$ ($|b_{1,N}|^2 \equiv b^2$), nonlocal transport is absent, while the other 
contributions vanish as 
$G^{LAR}_{j}\simeq 2 g^2/b^4$ and $G_j^B\simeq 2g/b^2$. This regime is 
reminiscent of Coulomb blockade in transport through normal quantum dots. In 
our case, instead of Coulomb blockade, we have a \textit{dissipative blockade} 
provided by the strong bath effect, $b^2\gg g$, and localization of the 
lowest-lying eigenstates at opposite ends of the chain.

\noindent\textsf{\color{blue} Numerical results.} To illustrate the above results, we consider the sweet spot of the Kitaev chain 
($\mu = 0$, $|t| = |\Delta|$), assume symmetric leads with $g_{1,N}=g$, and 
choose dissipative fields such that $\underline{\mu} \in \mathbb{R}^{N}$ and 
$\underline{\nu} \in i\mathbb{R}^{N}$. To highlight nonlocal transport, we 
further assume that $\underline{\mu}$ ($\underline{\nu}$) does not act on the 
right (left) end of the chain, i.e., $\mu^r_N=0$ ($\nu^i_1=0$). Since the MBS 
eigenfunctions are $\chi_{j,n}=\delta_{n,j}$, the vectors \eqref{sa_vecs} 
reduce to $\bm{b}_{1} = \{\mu_1^r,0\}/2$ and 
$\bm{b}_{N} = \{0,-\nu_{N}^i\}/2$, so that $\bm{b}_1\cdot \bm{b}_N = 0$ and 
$p_g=p=0$. 

With this parameter choice we use Eq.~\eqref{G_terms} to generate the ($\mu_{1}^{r}$, $\nu_{N}^{i}$)-maps of the 
left-lead conductance contributions at zero bias (see Figs.~\ref{G1_pt}a,b). 
Along the diagonals 
$|\mu_{1}^{r}|{=}|\nu_{N}^{i}|$ (which correspond to the symmetric case 
$|\bm{b}_{1,N}|=b$),  
$G_{1}^{EC}$ reaches 
$1/4$ for $|\mu_{1}^{r}|,|\nu_{N}^{i}|\gg g$, in agreement with the analysis 
above (see pink dots in Fig.~\ref{G1_pt}\,a). In contrast, $G_{1}^{EC}$ 
vanishes near the lines $\mu_{1}^{r} = 0$ and $\nu_{N}^{i} = 0$, since then 
$\bm{b}_N\times \bm{b}_1\equiv 0$.

For $|\mu_{1}^r|$ close to zero, dissipation on the left 
MM $\underline{\chi}_1$ vanishes,  
producing a quantized LAR 
contribution, $G_{1}^{LAR}\approx 2$ \cite{law-09} (see Fig.~\ref{G1_pt}b). 
This feature is insensitive to $|\nu_{N}^i|$, as the lowest-lying eigenstates 
remain localized at opposite ends of the chain. For $|\mu_{1}^r| \gg 
|\nu_{N}^i|$, the zero-bias peak is suppressed, $G_{1}^{LAR} \approx 0$, due 
to strong fermion loss~\cite{aksenov-26}. At the map's corners, 
$G_{1}^{LAR} \approx 1/2$, again consistent with our analysis. 
The bath-induced contribution has a peak $G_1^B \approx G_{1}^{LAR} \approx 1/2$ at $|\mu_{1}^r| \approx 2\sqrt{g}$ 
and $|\nu_{N}^i| \approx 0$.
For 
$|\mu_{1}^r|,|\nu_{N}^i| \gg g$, $G_1^B$ becomes negligible.

For the right-lead conductance at zero bias, $G_N$, the features near 
$|\mu_{1}^r| \approx 0$ or $|\nu_{N}^i| \approx 0$ reappear under the 
exchange $\mu_{1}^r \leftrightarrow \nu_{N}^i$. Moreover, in the symmetric 
lead case, a $\pi/2$ rotation of the maps in Figs.~\ref{G1_pt}\,a,b suffices 
to generate the corresponding maps for $G_N^{EC/LAR}$.

Bath-assisted control of the open MBS parity can  be achieved, e.g.,  with purely real 
dissipative fields, $\underline{\mu}, \underline{\nu} \in \mathbb{R}^{N}$. 
Assuming again that $\underline{\mu}$ ($\underline{\nu}$) does not act on the 
right (left) end ($\mu^r_N=0$, $\nu^r_1=0$), the vectors \eqref{sa_vecs} 
become $\bm{b}_{1} {=} \{\mu_1^r,0\}/2$ and 
$\bm{b}_{N} {=} \{-\nu_{N}^r,0\}/2$. Then $p_g$ can become nonzero since 
$\bm{b}_N \times \bm{b}_1 = 0$, and hence $G_j^{EC/CAR}=0$.
In the symmetric case 
$|b_{1,N}|^2=b^2\gg g$, the reduced density matrix \eqref{rho} approaches a 
pure state with $|p_g|\approx|p|=1$ (violet dots in Fig.~\ref{G1_pt}c). 
This realizes the dissipative blockade regime. Since $\bm{b}_N{\times}\bm{b}_1 = 0$ 
prevents hybridization of the MBS wave functions, varying $\nu_{N}^r$ does not 
affect $G_{1}^{LAR/B}$, but changes $p_g$ via Eq.~\eqref{rho}. Thus, for a fixed 
$\mu_1^r$, the conductance contributions are independent of $p_g$ (see 
Fig.~\ref{G1_pt}c), allowing the MBS parity to be tuned while the system 
response remains unchanged.
There is a direct correspondence between $p_g$ in the regime $G_1^{{EC/CAR}}{=}0$ and $G_1^{{EC/CAR}}$ in the regime $p_g=0$: $|p_g(\nu_N^r)| {\leftrightarrow} G_1(i\nu_N^r)$. This suggests a protocol for Majorana qubit readout: first fix the bath parity with real $\nu_N^r$, then change the phase of $\nu_N$ by $\pi/2$ (so that it becomes purely imaginary, with strictly zero parity), and measure the conductance, which encodes the original parity.

\noindent\textsf{\color{blue} Phase coherence.} A specific example of dissipative fields that maximize the nonlocal 
conductance contributions corresponds to the jump operator
$\hat{L} =  \mu^r_{1}(\hat{c}_1 {+} i\hat{c}^{\dagger}_{N}) + \nu^r_{1}(\hat{c}_1^\dagger {-} i\hat{c}_{N})$.
Then, at the sweet spot, the vectors \eqref{sa_vecs} 
become 
$\bm{b}_{1} {=} \{\mu_1^r{+}\nu^r_1,0\}/2$ and 
$\bm{b}_{N} {=} \{0,{-}\nu_{1}^r{-}\mu^r_1\}/2$. This structure 
suggests an interpretation in terms of coherent particle transfer between the 
first and last sites of the chain (see Fig.~\ref{model}). But is the nonlocal 
transport facilitated by such jump operator 
indeed phase coherent?

To demonstrate phase coherence explicitly, we consider a double-path setup 
(see Fig.~\ref{model}b). The 
upper arm is the Kitaev chain at the sweet 
spot with gain and loss described by the jump operator 
$\hat{L} =  \mu^r_{1}\hat{c}_1 + \nu_{N}\hat{c}^{\dagger}_{N}$. The 
lower arm is a normal chain of the same length $N$, serving as a 
reference arm, with the Hamiltonian
$H_0=\sum_{n=1}^N\xi_0\,\hat{a}_{n}^\dagger\hat{a}_n {+}\sum_{n=1}^{N-1} [t_0 \hat{a}_{n}^\dagger \hat{a}_{n+1} {+} {\rm h.c.}]$. 
Tunneling between the normal chain and the leads is described by 
$\hat{H}'_{T}= \sum_{k} [t_{1}'e^{i\phi/2}\hat{d}_{1k}^{\dagger}\hat{a}_{1}{+}t_{N}'e^{-i\phi/2}\hat{d}_{Nk}^{\dagger}\hat{a}_{N}{+} {\rm h.c.} ]$ (see End Matter). 
Here $\hat{a}_{n}$ ($\hat{a}_{n}^\dagger$) annihilates (creates) an electron on 
the $n$-th site of the reference arm with energy $\xi_{0}$, $t_{0}$ is the 
hopping amplitude, 
$t_{1,N}'$ are tunneling amplitudes 
between the leads and the nearest sites of the reference arm, and $\phi$ is 
the AB phase. 
We assume a symmetric lead 
configuration and weak tunneling to the leads, 
$|\bm{b}_{1,N}|^2=b^2\gg g_{1,N}=g$.

\begin{figure}[t]
\centerline{\includegraphics[width=0.5\textwidth]{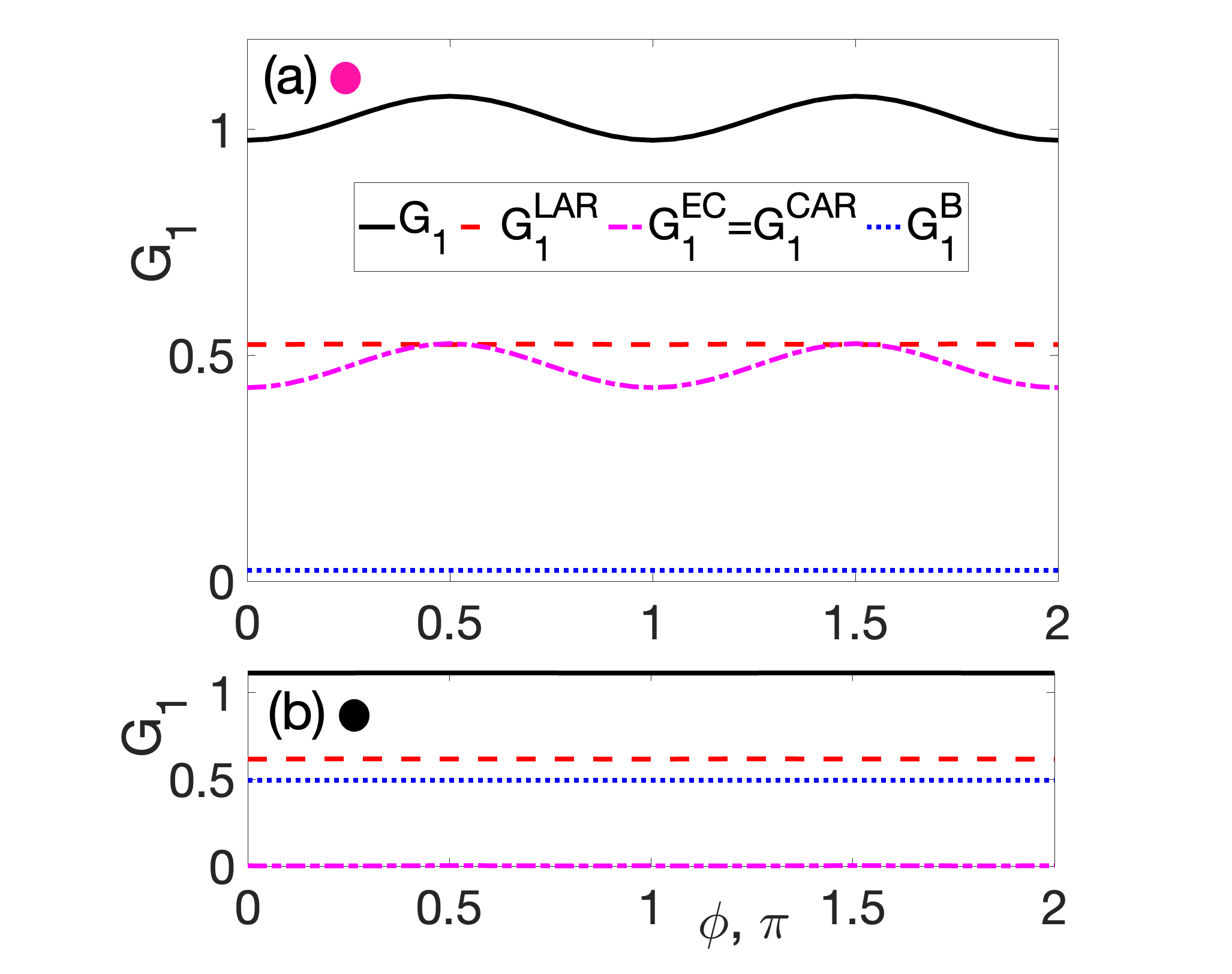}}
		\caption{The $\phi$-dependence of the contributions to the left-lead conductance (AB effect)
        in the linear-response regime and at the sweet spot. (a) the regime of maximal nonlocal conductance, corresponding to pink dots in Fig. \ref{G1_pt}a; 
        (b) the regime of absent nonlocal conductance,
        corresponding to black dot in Fig. \ref{G1_pt}c. 
        Parameters used: $g_{1,N}=g_{1,N}'=0.05$.}
\label{G1_AB}
\end{figure}

We have numerically studied the AB effect on the conductance of this 
double-arm system. Oscillations in the AB setup can only emerge if both arms 
participate coherently in nonlocal transport. The reference arm, being a 
normal-metal chain, provides a finite though generally small contribution to 
the total conductance, as it lacks low-energy excitations. Therefore, in the 
regime considered, AB oscillations of the total conductance serve as a 
hallmark of nonlocal CAR and EC contributions, while LAR and bath terms do 
not contribute to the interference pattern. This conclusion is supported by 
the plots in Fig.~\ref{G1_AB}a for the case of $\nu_N=i\mu_1^r$ (pink dots in Fig. \ref{G1_pt}a), 
for which $p=0$. 
Clear conductance oscillations as a function of flux $\phi$ 
appear in $G_1$ (see black curve in Fig.~\ref{G1_AB}a). Moreover, these 
oscillations are indeed entirely due to the nonlocal CAR and EC contributions 
(magenta dash-dotted curve), while LAR and bath terms show no interference 
(red dashed and blue dotted curves in Fig.~\ref{G1_AB}a). To further verify 
that AB oscillations vanish identically when
EC and CAR are absent, we consider purely real dissipative fields, i.e. $\nu_N=\mu_1^r\approx2\sqrt{g}$ 
(black dot in Fig. \ref{G1_pt}c). As shown in 
Fig.~\ref{G1_AB}b, no AB oscillations are visible, since the 
upper arm does 
not support nonlocal transport. 

\begin{figure}[t]
\centerline{\includegraphics[width=0.4\textwidth]{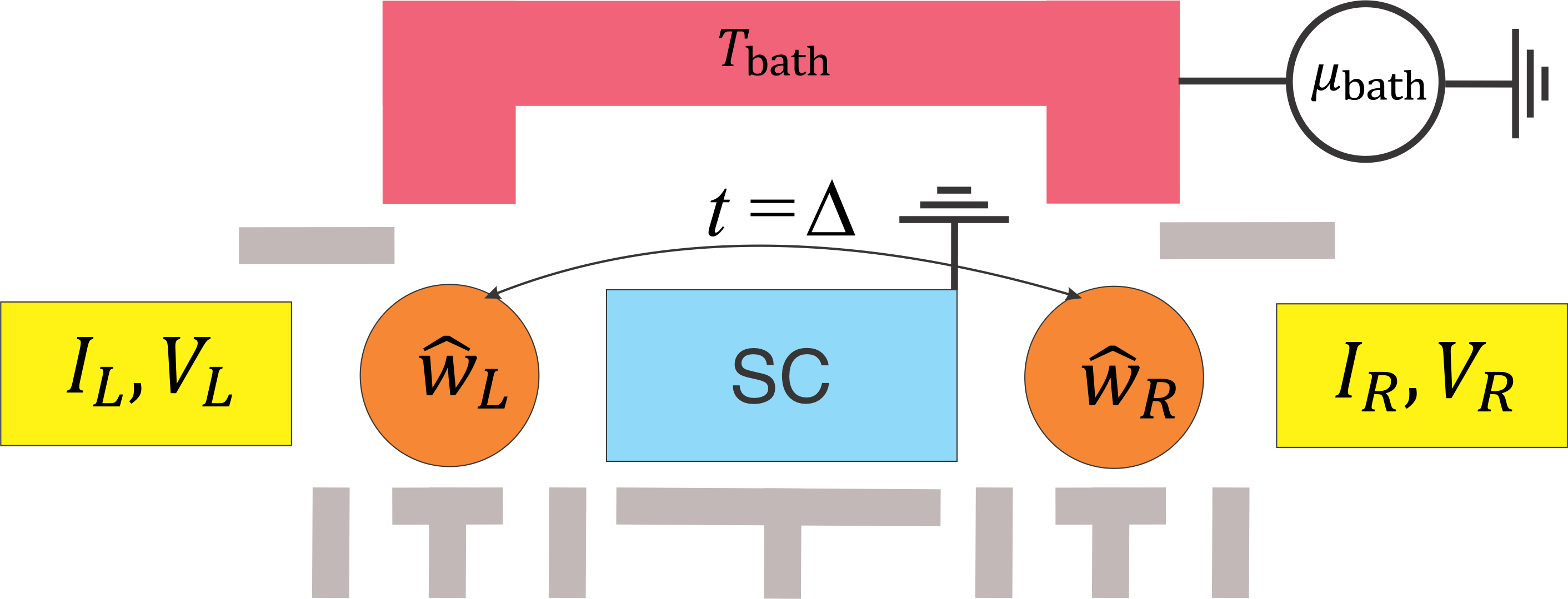}}
		\caption{Sketch of a possible experiment to detect nonlocal transport by dissipation (see text).}
\label{exptprop}
\end{figure}

\noindent\textsf{\color{blue} Discussions and conclusions.} Our analysis shows that in dissipative topological superconductors of class BDI, quantum transport can be controlled by tuning environmental parameters. 
Unlike the conventional dissipationless setup with a Majorana wire between two 
leads~\cite{bolech-07,law-09}, we find that 
EC and CAR
enable charge transfer between spatially separated regions 
where MMs of the isolated system were localized. This effect arises from bath-induced 
hybridization of MMs and should not be confused with their 
direct hybridization~\cite{nilsson-08,wu-12}. 

We also identified a dissipative blockade regime where the 
conductance is suppressed and the MBS parity is 
pinned near $\pm1$ (i.e., $\rho_{\rm red}$ is nearly pure). This behavior is analogous to a standard Coulomb blockade.

		The discussed effects can be detected in the system of superconducting quantum dots which is actively studied nowadays in the context of poor man's MBS and their use in quantum computations \cite{leijnse-12,tsintzis-22}. A minimal Kitaev chain consisting of two dots at the sweet spot has recently been implemented \cite{dvir-23,tenhaaf-24}. This is already a proper testbed to demonstrate the nonlocal phase-coherent transport by dissipation. 		In the latest experiments \cite{vanloo-26,zhang-26b,zatelli-26} interaction of the MBS with  quasiparticles is an ubiqutous problem affecting the chain's ground-state parity. Since it is difficult to
        control  the internal bath we suggest to supplement the 
        three-terminal setups \cite{dvir-23,tenhaaf-24} with the Fermi reservoir placed in the center and coupled with both dots (see Fig.~\ref{exptprop}). To guarantee the fast relaxation 
        to equilibrium state
        of this central lead one can control both its size and temperature \cite{pothier-97,giazotto-06}. For example, a wide-band metallic lead with the electrochemical potential $|\mu_{\rm bath}| \gg t$ or the temperature $T_{\rm bath} \gg t$ can be utilized \cite{breuer-02,dorda-17,jin-20}. In the experiment \cite{dvir-23} the effective hopping was estimated as $t \sim 100$ mK. 
		Essentially, the conductance measurement time in the proposed four-terminal 
setup must be shorter than the parity lifetime determined by quasiparticle 
poisoning in superconducting double-dot systems. The latter typically ranges 
from $1$ to $100$ ms \cite{dvir-23,zatelli-26}.

Nonlocal charge transport can also be realized in optical lattices with cold 
atoms. For instance, without unitary dynamics, $N-1$ baths with dissipative 
fields in the gapped subspace of an isolated wire can engineer MBS 
\cite{diehl-11}. An additional weak $N$-th bath acting on both ends then 
generates nonlocal currents, yielding $G_{j}^{EC}/G_{j}^{LAR}\simeq 1/2$ 
(see End Matter).

To summarize, we study Kitaev chain coupled to an external Markovian environment and 
identify a regime where nonlocal conductance contributions reach the local 
one, each amounting to $\approx 1/2$. In this regime, the reduced density matrix of the 
lowest-energy subspace is maximally mixed (its 
entanglement spectrum is  
twofold degenerate) reminiscent of the 
nontrivial phase of an isolated topological superconductor. This behavior 
contrasts with the mechanism of
teleportation in Ref.~\cite{fu-10}, where fermion parity is fixed. 
Phase coherence of the dissipation-induced nonlocal transport is confirmed by 
AB oscillations of the conductance. Thus, we predict an effect 
resembling quantum teleportation of carriers between localized MMs 
mediated by an external Markovian environment. The correlation between nonlocal transport and the internal structure of the lowest-energy modes makes controlled dissipation a promising tool both for 
addressing the Majorana qubit and for reading out its state via differential 
conductance measurements.

\noindent\textsf{\color{blue} Acknowledgments.} This work was supported by Russian Science Foundation  under Grant No. \href{https://rscf.ru/project/26-12-00052/}{26-12-00052}. The authors acknowledge personal support from the Foundation for the Advancement of Theoretical Physics and Mathematics ``BASIS''.

\noindent\textsf{\color{blue} Data Availability.} The data that support the findings of
this article are not publicly available. The data are available from the authors upon reasonable request.
    
\bibliography{Majorana}

\providecommand{\noopsort}[1]{}\providecommand{\singleletter}[1]{#1}%
\begin{thebibliography}{65}%
\makeatletter
\providecommand \@ifxundefined [1]{%
 \@ifx{#1\undefined}
}%
\providecommand \@ifnum [1]{%
 \ifnum #1\expandafter \@firstoftwo
 \else \expandafter \@secondoftwo
 \fi
}%
\providecommand \@ifx [1]{%
 \ifx #1\expandafter \@firstoftwo
 \else \expandafter \@secondoftwo
 \fi
}%
\providecommand \natexlab [1]{#1}%
\providecommand \enquote  [1]{``#1''}%
\providecommand \bibnamefont  [1]{#1}%
\providecommand \bibfnamefont [1]{#1}%
\providecommand \citenamefont [1]{#1}%
\providecommand \href@noop [0]{\@secondoftwo}%
\providecommand \href [0]{\begingroup \@sanitize@url \@href}%
\providecommand \@href[1]{\@@startlink{#1}\@@href}%
\providecommand \@@href[1]{\endgroup#1\@@endlink}%
\providecommand \@sanitize@url [0]{\catcode `\\12\catcode `\$12\catcode
  `\&12\catcode `\#12\catcode `\^12\catcode `\_12\catcode `\%12\relax}%
\providecommand \@@startlink[1]{}%
\providecommand \@@endlink[0]{}%
\providecommand \url  [0]{\begingroup\@sanitize@url \@url }%
\providecommand \@url [1]{\endgroup\@href {#1}{\urlprefix }}%
\providecommand \urlprefix  [0]{URL }%
\providecommand \Eprint [0]{\href }%
\providecommand \doibase [0]{https://doi.org/}%
\providecommand \selectlanguage [0]{\@gobble}%
\providecommand \bibinfo  [0]{\@secondoftwo}%
\providecommand \bibfield  [0]{\@secondoftwo}%
\providecommand \translation [1]{[#1]}%
\providecommand \BibitemOpen [0]{}%
\providecommand \bibitemStop [0]{}%
\providecommand \bibitemNoStop [0]{.\EOS\space}%
\providecommand \EOS [0]{\spacefactor3000\relax}%
\providecommand \BibitemShut  [1]{\csname bibitem#1\endcsname}%
\let\auto@bib@innerbib\@empty
\bibitem [{\citenamefont {Moore}\ and\ \citenamefont {Read}(1991)}]{moore-91}%
  \BibitemOpen
  \bibfield  {author} {\bibinfo {author} {\bibfnamefont {G.}~\bibnamefont
  {Moore}}\ and\ \bibinfo {author} {\bibfnamefont {N.}~\bibnamefont {Read}},\
  }\bibfield  {title} {\bibinfo {title} {Nonabelions in the fractional quantum
  {H}all effect},\ }\href {https://doi.org/10.1016/0550-3213(91)90407-O}
  {\bibfield  {journal} {\bibinfo  {journal} {Nucl.\ Phys. B}\ }\textbf
  {\bibinfo {volume} {360}},\ \bibinfo {pages} {362} (\bibinfo {year}
  {1991})}\BibitemShut {NoStop}%
\bibitem [{\citenamefont {Volovik}(1999)}]{volovik-99}%
  \BibitemOpen
  \bibfield  {author} {\bibinfo {author} {\bibfnamefont {G.~E.}\ \bibnamefont
  {Volovik}},\ }\bibfield  {title} {\bibinfo {title} {Fermion zero modes on
  vortices in chiral superconductors},\ }\href
  {https://doi.org/10.1134/1.568223} {\bibfield  {journal} {\bibinfo  {journal}
  {JETP Lett.}\ }\textbf {\bibinfo {volume} {70}},\ \bibinfo {pages} {609}
  (\bibinfo {year} {1999})}\BibitemShut {NoStop}%
\bibitem [{\citenamefont {Fu}\ and\ \citenamefont {Kane}(2008)}]{fu-08}%
  \BibitemOpen
  \bibfield  {author} {\bibinfo {author} {\bibfnamefont {L.}~\bibnamefont
  {Fu}}\ and\ \bibinfo {author} {\bibfnamefont {C.~L.}\ \bibnamefont {Kane}},\
  }\bibfield  {title} {\bibinfo {title} {Superconducting proximity effect and
  {M}ajorana fermions at the surface of a topological insulator},\ }\href
  {https://doi.org/10.1016/0550-3213(91)90407-O} {\bibfield  {journal}
  {\bibinfo  {journal} {Phys.\ Rev.\ Lett.}\ }\textbf {\bibinfo {volume}
  {100}},\ \bibinfo {pages} {096407} (\bibinfo {year} {2008})}\BibitemShut
  {NoStop}%
\bibitem [{\citenamefont {Lutchyn}\ \emph {et~al.}(2010)\citenamefont
  {Lutchyn}, \citenamefont {Sau},\ and\ \citenamefont {Sarma}}]{lutchyn-10}%
  \BibitemOpen
  \bibfield  {author} {\bibinfo {author} {\bibfnamefont {R.~M.}\ \bibnamefont
  {Lutchyn}}, \bibinfo {author} {\bibfnamefont {J.~D.}\ \bibnamefont {Sau}},\
  and\ \bibinfo {author} {\bibfnamefont {S.~D.}\ \bibnamefont {Sarma}},\
  }\bibfield  {title} {\bibinfo {title} {{M}ajorana fermions and a topological
  phase transition in semiconductor-superconductor heterostructures},\ }\href
  {https://doi.org/10.1103/PhysRevLett.105.077001} {\bibfield  {journal}
  {\bibinfo  {journal} {Phys.\ Rev.\ Lett.}\ }\textbf {\bibinfo {volume}
  {105}},\ \bibinfo {pages} {077001} (\bibinfo {year} {2010})}\BibitemShut
  {NoStop}%
\bibitem [{\citenamefont {Oreg}\ \emph {et~al.}(2010)\citenamefont {Oreg},
  \citenamefont {Refael},\ and\ \citenamefont {von Oppen}}]{oreg-10}%
  \BibitemOpen
  \bibfield  {author} {\bibinfo {author} {\bibfnamefont {Y.}~\bibnamefont
  {Oreg}}, \bibinfo {author} {\bibfnamefont {G.}~\bibnamefont {Refael}},\ and\
  \bibinfo {author} {\bibfnamefont {F.}~\bibnamefont {von Oppen}},\ }\bibfield
  {title} {\bibinfo {title} {Helical liquids and {M}ajorana bound states in
  quantum wires},\ }\href {https://doi.org/10.1103/PhysRevLett.105.177002}
  {\bibfield  {journal} {\bibinfo  {journal} {Phys.\ Rev.\ Lett.}\ }\textbf
  {\bibinfo {volume} {105}},\ \bibinfo {pages} {177002} (\bibinfo {year}
  {2010})}\BibitemShut {NoStop}%
\bibitem [{\citenamefont {Nadj-Perge}\ \emph {et~al.}(2013)\citenamefont
  {Nadj-Perge}, \citenamefont {Drozdov}, \citenamefont {Bernevig},\ and\
  \citenamefont {Yazdani}}]{nadj-perge-13}%
  \BibitemOpen
  \bibfield  {author} {\bibinfo {author} {\bibfnamefont {S.}~\bibnamefont
  {Nadj-Perge}}, \bibinfo {author} {\bibfnamefont {I.~K.}\ \bibnamefont
  {Drozdov}}, \bibinfo {author} {\bibfnamefont {B.~A.}\ \bibnamefont
  {Bernevig}},\ and\ \bibinfo {author} {\bibfnamefont {A.}~\bibnamefont
  {Yazdani}},\ }\bibfield  {title} {\bibinfo {title} {Proposal for realizing
  {M}ajorana fermions in chains of magnetic atoms on a superconductor},\ }\href
  {https://doi.org/10.1103/PhysRevB.88.020407} {\bibfield  {journal} {\bibinfo
  {journal} {Phys.\ Rev.\ B}\ }\textbf {\bibinfo {volume} {88}},\ \bibinfo
  {pages} {020407} (\bibinfo {year} {2013})}\BibitemShut {NoStop}%
\bibitem [{\citenamefont {Backens}\ \emph {et~al.}(2022)\citenamefont
  {Backens}, \citenamefont {Shnirman},\ and\ \citenamefont
  {Makhlin}}]{backens-22}%
  \BibitemOpen
  \bibfield  {author} {\bibinfo {author} {\bibfnamefont {S.}~\bibnamefont
  {Backens}}, \bibinfo {author} {\bibfnamefont {A.}~\bibnamefont {Shnirman}},\
  and\ \bibinfo {author} {\bibfnamefont {Y.}~\bibnamefont {Makhlin}},\
  }\bibfield  {title} {\bibinfo {title} {Topological {J}osephson junction in
  transverse magnetic field},\ }\href
  {https://doi.org/10.1134%2FS0021364022602561} {\bibfield  {journal} {\bibinfo
   {journal} {JETP Lett.}\ }\textbf {\bibinfo {volume} {116}},\ \bibinfo
  {pages} {891} (\bibinfo {year} {2022})}\BibitemShut {NoStop}%
\bibitem [{\citenamefont {Yazdani}\ \emph {et~al.}(2023)\citenamefont
  {Yazdani}, \citenamefont {von Oppen}, \citenamefont {Halperin},\ and\
  \citenamefont {Yacoby}}]{yazdani-23}%
  \BibitemOpen
  \bibfield  {author} {\bibinfo {author} {\bibfnamefont {A.}~\bibnamefont
  {Yazdani}}, \bibinfo {author} {\bibfnamefont {F.}~\bibnamefont {von Oppen}},
  \bibinfo {author} {\bibfnamefont {B.}~\bibnamefont {Halperin}},\ and\
  \bibinfo {author} {\bibfnamefont {A.}~\bibnamefont {Yacoby}},\ }\bibfield
  {title} {\bibinfo {title} {Hunting for {M}ajoranas},\ }\href
  {https://doi.org/10.1126/science.ade0850} {\bibfield  {journal} {\bibinfo
  {journal} {Science}\ }\textbf {\bibinfo {volume} {380}},\ \bibinfo {pages}
  {6651} (\bibinfo {year} {2023})}\BibitemShut {NoStop}%
\bibitem [{\citenamefont {Sarma}(2023)}]{dassarma-23}%
  \BibitemOpen
  \bibfield  {author} {\bibinfo {author} {\bibfnamefont {S.~D.}\ \bibnamefont
  {Sarma}},\ }\bibfield  {title} {\bibinfo {title} {In search of {M}ajorana},\
  }\href {https://doi.org/10.1038/s41567-022-01900-9} {\bibfield  {journal}
  {\bibinfo  {journal} {Nat. Phys.}\ }\textbf {\bibinfo {volume} {19}},\
  \bibinfo {pages} {165} (\bibinfo {year} {2023})}\BibitemShut {NoStop}%
\bibitem [{\citenamefont {Frolov}\ \emph {et~al.}(2023)\citenamefont {Frolov},
  \citenamefont {Zhang}, \citenamefont {Zhang}, \citenamefont {Jiang},
  \citenamefont {Byard}, \citenamefont {Mudi}, \citenamefont {Chen},
  \citenamefont {Chen}, \citenamefont {Hocevar}, \citenamefont {Gupta},
  \citenamefont {Riggert},\ and\ \citenamefont {Pribiag}}]{frolov-23}%
  \BibitemOpen
  \bibfield  {author} {\bibinfo {author} {\bibfnamefont {S.}~\bibnamefont
  {Frolov}}, \bibinfo {author} {\bibfnamefont {P.}~\bibnamefont {Zhang}},
  \bibinfo {author} {\bibfnamefont {B.}~\bibnamefont {Zhang}}, \bibinfo
  {author} {\bibfnamefont {Y.}~\bibnamefont {Jiang}}, \bibinfo {author}
  {\bibfnamefont {S.}~\bibnamefont {Byard}}, \bibinfo {author} {\bibfnamefont
  {S.}~\bibnamefont {Mudi}}, \bibinfo {author} {\bibfnamefont {J.}~\bibnamefont
  {Chen}}, \bibinfo {author} {\bibfnamefont {A.-H.}\ \bibnamefont {Chen}},
  \bibinfo {author} {\bibfnamefont {M.}~\bibnamefont {Hocevar}}, \bibinfo
  {author} {\bibfnamefont {M.}~\bibnamefont {Gupta}}, \bibinfo {author}
  {\bibfnamefont {C.}~\bibnamefont {Riggert}},\ and\ \bibinfo {author}
  {\bibfnamefont {V.}~\bibnamefont {Pribiag}},\ }\href@noop {} {\bibinfo
  {title} {''{S}moking gun'' signatures of topological milestones in trivial
  materials by measurement fine-tuning and data postselection}} (\bibinfo
  {year} {2023}),\ \Eprint {https://arxiv.org/abs/arXiv:2309.09368}
  {arXiv:2309.09368} \BibitemShut {NoStop}%
\bibitem [{\citenamefont {Legg}(2026)}]{legg-26}%
  \BibitemOpen
  \bibfield  {author} {\bibinfo {author} {\bibfnamefont {H.~F.}\ \bibnamefont
  {Legg}},\ }\bibfield  {title} {\bibinfo {title} {On the robustness of
  topological gap detection via transport},\ }\href
  {https://doi.org/10.1038/s41586-026-10567-8} {\bibfield  {journal} {\bibinfo
  {journal} {Nature}\ }\textbf {\bibinfo {volume} {654}},\ \bibinfo {pages}
  {E22} (\bibinfo {year} {2026})}\BibitemShut {NoStop}%
\bibitem [{\citenamefont {{Microsoft Quantum}}(2026)}]{microsoft-26}%
  \BibitemOpen
  \bibfield  {author} {\bibinfo {author} {\bibnamefont {{Microsoft Quantum}}},\
  }\bibfield  {title} {\bibinfo {title} {Reply to: On the robustness of
  topological gap detection via transport},\ }\href
  {https://doi.org/10.1038/s41586-026-10568-7} {\bibfield  {journal} {\bibinfo
  {journal} {Nature}\ }\textbf {\bibinfo {volume} {654}},\ \bibinfo {pages}
  {E27} (\bibinfo {year} {2026})}\BibitemShut {NoStop}%
\bibitem [{\citenamefont {Kitaev}(2001)}]{kitaev-01}%
  \BibitemOpen
  \bibfield  {author} {\bibinfo {author} {\bibfnamefont {A.~Y.}\ \bibnamefont
  {Kitaev}},\ }\bibfield  {title} {\bibinfo {title} {Unpaired {M}ajorana
  fermions in quantum wires},\ }\href
  {https://doi.org/http://dx.doi.org/10.1070/1063-7869/44/10S/S29} {\bibfield
  {journal} {\bibinfo  {journal} {Phys.\ Usp.}\ }\textbf {\bibinfo {volume}
  {44}},\ \bibinfo {pages} {131} (\bibinfo {year} {2001})}\BibitemShut
  {NoStop}%
\bibitem [{\citenamefont {Leijnse}\ and\ \citenamefont
  {Flensberg}(2012)}]{leijnse-12}%
  \BibitemOpen
  \bibfield  {author} {\bibinfo {author} {\bibfnamefont {M.}~\bibnamefont
  {Leijnse}}\ and\ \bibinfo {author} {\bibfnamefont {K.}~\bibnamefont
  {Flensberg}},\ }\bibfield  {title} {\bibinfo {title} {Parity qubits and poor
  man's {M}ajorana bound states in double quantum dots},\ }\href
  {https://doi.org/10.1103/PhysRevB.86.134528} {\bibfield  {journal} {\bibinfo
  {journal} {Phys. Rev. B}\ }\textbf {\bibinfo {volume} {86}},\ \bibinfo
  {pages} {134528} (\bibinfo {year} {2012})}\BibitemShut {NoStop}%
\bibitem [{\citenamefont {Sau}\ and\ \citenamefont {Sarma}(2012)}]{sau-12}%
  \BibitemOpen
  \bibfield  {author} {\bibinfo {author} {\bibfnamefont {J.~D.}\ \bibnamefont
  {Sau}}\ and\ \bibinfo {author} {\bibfnamefont {S.~D.}\ \bibnamefont
  {Sarma}},\ }\bibfield  {title} {\bibinfo {title} {Realizing a robust
  practical {M}ajorana chain in a quantum-dot- superconductor linear array},\
  }\href {https://doi.org/10.1038/ncomms1966} {\bibfield  {journal} {\bibinfo
  {journal} {Nat.\ Comm.}\ }\textbf {\bibinfo {volume} {3}},\ \bibinfo {pages}
  {964} (\bibinfo {year} {2012})}\BibitemShut {NoStop}%
\bibitem [{\citenamefont {Dvir}\ \emph {et~al.}(2023)\citenamefont {Dvir},
  \citenamefont {Wang}, \citenamefont {van Loo}, \citenamefont {Liu},
  \citenamefont {Mazur}, \citenamefont {Bordin}, \citenamefont {ten Haaf},
  \citenamefont {Wang}, \citenamefont {van Driel}, \citenamefont {Zatelli},
  \citenamefont {Li}, \citenamefont {Malinowski}, \citenamefont {Gazibegovic},
  \citenamefont {Badawy}, \citenamefont {Bakkers}, \citenamefont {Wimmer},\
  and\ \citenamefont {Kouwenhoven}}]{dvir-23}%
  \BibitemOpen
  \bibfield  {author} {\bibinfo {author} {\bibfnamefont {T.}~\bibnamefont
  {Dvir}}, \bibinfo {author} {\bibfnamefont {G.}~\bibnamefont {Wang}}, \bibinfo
  {author} {\bibfnamefont {N.}~\bibnamefont {van Loo}}, \bibinfo {author}
  {\bibfnamefont {C.-X.}\ \bibnamefont {Liu}}, \bibinfo {author} {\bibfnamefont
  {G.~P.}\ \bibnamefont {Mazur}}, \bibinfo {author} {\bibfnamefont
  {A.}~\bibnamefont {Bordin}}, \bibinfo {author} {\bibfnamefont {S.~L.~D.}\
  \bibnamefont {ten Haaf}}, \bibinfo {author} {\bibfnamefont {J.-Y.}\
  \bibnamefont {Wang}}, \bibinfo {author} {\bibfnamefont {D.}~\bibnamefont {van
  Driel}}, \bibinfo {author} {\bibfnamefont {F.}~\bibnamefont {Zatelli}},
  \bibinfo {author} {\bibfnamefont {X.}~\bibnamefont {Li}}, \bibinfo {author}
  {\bibfnamefont {F.~K.}\ \bibnamefont {Malinowski}}, \bibinfo {author}
  {\bibfnamefont {S.}~\bibnamefont {Gazibegovic}}, \bibinfo {author}
  {\bibfnamefont {G.}~\bibnamefont {Badawy}}, \bibinfo {author} {\bibfnamefont
  {E.~P. A.~M.}\ \bibnamefont {Bakkers}}, \bibinfo {author} {\bibfnamefont
  {M.}~\bibnamefont {Wimmer}},\ and\ \bibinfo {author} {\bibfnamefont {L.~P.}\
  \bibnamefont {Kouwenhoven}},\ }\bibfield  {title} {\bibinfo {title}
  {Realization of a minimal {K}itaev chain in coupled quantum dots},\ }\href
  {https://doi.org/10.1038/s41586-022-05585-1} {\bibfield  {journal} {\bibinfo
  {journal} {Nature}\ }\textbf {\bibinfo {volume} {614}},\ \bibinfo {pages}
  {445} (\bibinfo {year} {2023})}\BibitemShut {NoStop}%
\bibitem [{\citenamefont {ten Haaf}\ \emph {et~al.}(2024)\citenamefont {ten
  Haaf}, \citenamefont {Wang}, \citenamefont {Bozkurt}, \citenamefont {Liu},
  \citenamefont {Kulesh}, \citenamefont {Kim}, \citenamefont {Xiao},
  \citenamefont {Thomas}, \citenamefont {Manfra}, \citenamefont {Dvir},
  \citenamefont {Wimmer},\ and\ \citenamefont {Goswami}}]{tenhaaf-24}%
  \BibitemOpen
  \bibfield  {author} {\bibinfo {author} {\bibfnamefont {S.~L.~D.}\
  \bibnamefont {ten Haaf}}, \bibinfo {author} {\bibfnamefont {Q.}~\bibnamefont
  {Wang}}, \bibinfo {author} {\bibfnamefont {A.~M.}\ \bibnamefont {Bozkurt}},
  \bibinfo {author} {\bibfnamefont {C.-X.}\ \bibnamefont {Liu}}, \bibinfo
  {author} {\bibfnamefont {I.}~\bibnamefont {Kulesh}}, \bibinfo {author}
  {\bibfnamefont {P.}~\bibnamefont {Kim}}, \bibinfo {author} {\bibfnamefont
  {D.}~\bibnamefont {Xiao}}, \bibinfo {author} {\bibfnamefont {C.}~\bibnamefont
  {Thomas}}, \bibinfo {author} {\bibfnamefont {M.~J.}\ \bibnamefont {Manfra}},
  \bibinfo {author} {\bibfnamefont {T.}~\bibnamefont {Dvir}}, \bibinfo {author}
  {\bibfnamefont {M.}~\bibnamefont {Wimmer}},\ and\ \bibinfo {author}
  {\bibfnamefont {S.}~\bibnamefont {Goswami}},\ }\bibfield  {title} {\bibinfo
  {title} {A two-site {K}itaev chain in a two-dimensional electron gas},\
  }\href {https://doi.org/10.1038/s41586-024-07434-9} {\bibfield  {journal}
  {\bibinfo  {journal} {Nature}\ }\textbf {\bibinfo {volume} {630}},\ \bibinfo
  {pages} {329} (\bibinfo {year} {2024})}\BibitemShut {NoStop}%
\bibitem [{\citenamefont {van Loo}\ \emph {et~al.}(2026)\citenamefont {van
  Loo}, \citenamefont {Zatelli}, \citenamefont {Steffensen}, \citenamefont
  {Roovers}, \citenamefont {Wang}, \citenamefont {Caekenberghe}, \citenamefont
  {Bordin}, \citenamefont {van Driel}, \citenamefont {Zhang}, \citenamefont
  {Huisman}, \citenamefont {Badawy}, \citenamefont {Bakkers}, \citenamefont
  {Mazur}, \citenamefont {Aguado},\ and\ \citenamefont
  {Kouwenhoven}}]{vanloo-26}%
  \BibitemOpen
  \bibfield  {author} {\bibinfo {author} {\bibfnamefont {N.}~\bibnamefont {van
  Loo}}, \bibinfo {author} {\bibfnamefont {F.}~\bibnamefont {Zatelli}},
  \bibinfo {author} {\bibfnamefont {G.~O.}\ \bibnamefont {Steffensen}},
  \bibinfo {author} {\bibfnamefont {B.}~\bibnamefont {Roovers}}, \bibinfo
  {author} {\bibfnamefont {G.}~\bibnamefont {Wang}}, \bibinfo {author}
  {\bibfnamefont {T.~V.}\ \bibnamefont {Caekenberghe}}, \bibinfo {author}
  {\bibfnamefont {A.}~\bibnamefont {Bordin}}, \bibinfo {author} {\bibfnamefont
  {D.}~\bibnamefont {van Driel}}, \bibinfo {author} {\bibfnamefont
  {Y.}~\bibnamefont {Zhang}}, \bibinfo {author} {\bibfnamefont {W.~D.}\
  \bibnamefont {Huisman}}, \bibinfo {author} {\bibfnamefont {G.}~\bibnamefont
  {Badawy}}, \bibinfo {author} {\bibfnamefont {E.~P. A.~M.}\ \bibnamefont
  {Bakkers}}, \bibinfo {author} {\bibfnamefont {G.~P.}\ \bibnamefont {Mazur}},
  \bibinfo {author} {\bibfnamefont {R.}~\bibnamefont {Aguado}},\ and\ \bibinfo
  {author} {\bibfnamefont {L.~P.}\ \bibnamefont {Kouwenhoven}},\ }\bibfield
  {title} {\bibinfo {title} {Single-shot parity readout of a minimal {K}itaev
  chain},\ }\href {https://doi.org/10.1038/s41586-025-09927-7} {\bibfield
  {journal} {\bibinfo  {journal} {Nature}\ }\textbf {\bibinfo {volume} {650}},\
  \bibinfo {pages} {334} (\bibinfo {year} {2026})}\BibitemShut {NoStop}%
\bibitem [{\citenamefont {Zhang}\ \emph {et~al.}(2026)\citenamefont {Zhang},
  \citenamefont {Kulesh}, \citenamefont {ten Haaf}, \citenamefont {van Loo},
  \citenamefont {Zatelli}, \citenamefont {Degroote}, \citenamefont {Prosko},\
  and\ \citenamefont {Goswami}}]{zhang-26b}%
  \BibitemOpen
  \bibfield  {author} {\bibinfo {author} {\bibfnamefont {Y.}~\bibnamefont
  {Zhang}}, \bibinfo {author} {\bibfnamefont {I.}~\bibnamefont {Kulesh}},
  \bibinfo {author} {\bibfnamefont {S.~L.}\ \bibnamefont {ten Haaf}}, \bibinfo
  {author} {\bibfnamefont {N.}~\bibnamefont {van Loo}}, \bibinfo {author}
  {\bibfnamefont {F.}~\bibnamefont {Zatelli}}, \bibinfo {author} {\bibfnamefont
  {T.}~\bibnamefont {Degroote}}, \bibinfo {author} {\bibfnamefont {C.~G.}\
  \bibnamefont {Prosko}},\ and\ \bibinfo {author} {\bibfnamefont
  {S.}~\bibnamefont {Goswami}},\ }\bibfield  {title} {\bibinfo {title} {Gate
  reflectometry in a minimal {K}itaev chain device},\ }\href
  {https://doi.org/10.1103/7367-wbp9} {\bibfield  {journal} {\bibinfo
  {journal} {PRX Quantum}\ }\textbf {\bibinfo {volume} {7}},\ \bibinfo {pages}
  {020317} (\bibinfo {year} {2026})}\BibitemShut {NoStop}%
\bibitem [{\citenamefont {Tsintzis}\ \emph
  {et~al.}(2022{\natexlab{a}})\citenamefont {Tsintzis}, \citenamefont {Souto},
  \citenamefont {Flensberg}, \citenamefont {Danon},\ and\ \citenamefont
  {Leijnse}}]{tsintzis-24}%
  \BibitemOpen
  \bibfield  {author} {\bibinfo {author} {\bibfnamefont {A.}~\bibnamefont
  {Tsintzis}}, \bibinfo {author} {\bibfnamefont {R.~S.}\ \bibnamefont {Souto}},
  \bibinfo {author} {\bibfnamefont {K.}~\bibnamefont {Flensberg}}, \bibinfo
  {author} {\bibfnamefont {J.}~\bibnamefont {Danon}},\ and\ \bibinfo {author}
  {\bibfnamefont {M.}~\bibnamefont {Leijnse}},\ }\bibfield  {title} {\bibinfo
  {title} {{M}ajorana qubits and non-{A}belian physics in quantum dot–based
  minimal {K}itaev chains},\ }\href
  {https://doi.org/10.1103/PRXQuantum.5.010323} {\bibfield  {journal} {\bibinfo
   {journal} {Phys. Rev. X}\ }\textbf {\bibinfo {volume} {5}},\ \bibinfo
  {pages} {010323} (\bibinfo {year} {2022}{\natexlab{a}})}\BibitemShut
  {NoStop}%
\bibitem [{\citenamefont {Pino}\ \emph {et~al.}(2024)\citenamefont {Pino},
  \citenamefont {Souto},\ and\ \citenamefont {Aguado}}]{pino-24}%
  \BibitemOpen
  \bibfield  {author} {\bibinfo {author} {\bibfnamefont {D.~M.}\ \bibnamefont
  {Pino}}, \bibinfo {author} {\bibfnamefont {R.~S.}\ \bibnamefont {Souto}},\
  and\ \bibinfo {author} {\bibfnamefont {R.}~\bibnamefont {Aguado}},\
  }\bibfield  {title} {\bibinfo {title} {Minimal {K}itaev-transmon qubit based
  on double quantum dots},\ }\href
  {https://doi.org/10.1103/PhysRevB.109.075101} {\bibfield  {journal} {\bibinfo
   {journal} {Phys. Rev. B}\ }\textbf {\bibinfo {volume} {109}},\ \bibinfo
  {pages} {075101} (\bibinfo {year} {2024})}\BibitemShut {NoStop}%
\bibitem [{\citenamefont {Zatelli}\ \emph {et~al.}(2026)\citenamefont
  {Zatelli}, \citenamefont {Roovers}, \citenamefont {van Loo}, \citenamefont
  {Lombardi}, \citenamefont {Luna}, \citenamefont {Miles}, \citenamefont
  {Sietses}, \citenamefont {Evertsz'}, \citenamefont {Fariña}, \citenamefont
  {Bordin}, \citenamefont {Badawy}, \citenamefont {Bakkers}, \citenamefont
  {Wimmer},\ and\ \citenamefont {Kouwenhoven}}]{zatelli-26}%
  \BibitemOpen
  \bibfield  {author} {\bibinfo {author} {\bibfnamefont {F.}~\bibnamefont
  {Zatelli}}, \bibinfo {author} {\bibfnamefont {B.}~\bibnamefont {Roovers}},
  \bibinfo {author} {\bibfnamefont {N.}~\bibnamefont {van Loo}}, \bibinfo
  {author} {\bibfnamefont {A.}~\bibnamefont {Lombardi}}, \bibinfo {author}
  {\bibfnamefont {J.~D.~T.}\ \bibnamefont {Luna}}, \bibinfo {author}
  {\bibfnamefont {S.}~\bibnamefont {Miles}}, \bibinfo {author} {\bibfnamefont
  {V.~P.}\ \bibnamefont {Sietses}}, \bibinfo {author} {\bibfnamefont
  {F.~J.~B.}\ \bibnamefont {Evertsz'}}, \bibinfo {author} {\bibfnamefont
  {P.~C.}\ \bibnamefont {Fariña}}, \bibinfo {author} {\bibfnamefont
  {A.}~\bibnamefont {Bordin}}, \bibinfo {author} {\bibfnamefont
  {G.}~\bibnamefont {Badawy}}, \bibinfo {author} {\bibfnamefont {E.~P.}\
  \bibnamefont {Bakkers}}, \bibinfo {author} {\bibfnamefont {M.}~\bibnamefont
  {Wimmer}},\ and\ \bibinfo {author} {\bibfnamefont {L.~P.}\ \bibnamefont
  {Kouwenhoven}},\ }\href {https://doi.org/10.48550/arXiv.2607.09511} {\bibinfo
  {title} {{M}ajorana parity qubit in coupled minimal {K}itaev chains}}
  (\bibinfo {year} {2026}),\ \Eprint {https://arxiv.org/abs/arXiv:2607.09511}
  {arXiv:2607.09511} \BibitemShut {NoStop}%
\bibitem [{\citenamefont {Fu}(2010)}]{fu-10}%
  \BibitemOpen
  \bibfield  {author} {\bibinfo {author} {\bibfnamefont {L.}~\bibnamefont
  {Fu}},\ }\bibfield  {title} {\bibinfo {title} {Electron teleportation via
  {M}ajorana bound states in a mesoscopic superconductor},\ }\href
  {https://doi.org/10.1103/PhysRevLett.104.056402} {\bibfield  {journal}
  {\bibinfo  {journal} {Phys.\ Rev.\ Lett.}\ }\textbf {\bibinfo {volume}
  {104}},\ \bibinfo {pages} {056402} (\bibinfo {year} {2010})}\BibitemShut
  {NoStop}%
\bibitem [{\citenamefont {Nitsch}\ \emph {et~al.}(2025)\citenamefont {Nitsch},
  \citenamefont {Maffi}, \citenamefont {Baran}, \citenamefont {Seoane~Souto},
  \citenamefont {Paaske}, \citenamefont {Leijnse},\ and\ \citenamefont
  {Burrello}}]{nitsch-25}%
  \BibitemOpen
  \bibfield  {author} {\bibinfo {author} {\bibfnamefont {M.}~\bibnamefont
  {Nitsch}}, \bibinfo {author} {\bibfnamefont {L.}~\bibnamefont {Maffi}},
  \bibinfo {author} {\bibfnamefont {V.~V.}\ \bibnamefont {Baran}}, \bibinfo
  {author} {\bibfnamefont {R.}~\bibnamefont {Seoane~Souto}}, \bibinfo {author}
  {\bibfnamefont {J.}~\bibnamefont {Paaske}}, \bibinfo {author} {\bibfnamefont
  {M.}~\bibnamefont {Leijnse}},\ and\ \bibinfo {author} {\bibfnamefont
  {M.}~\bibnamefont {Burrello}},\ }\bibfield  {title} {\bibinfo {title} {Poor
  man’s {M}ajorana tetron},\ }\href {https://doi.org/10.1103/r75t-jv32}
  {\bibfield  {journal} {\bibinfo  {journal} {PRX Quantum}\ }\textbf {\bibinfo
  {volume} {6}},\ \bibinfo {pages} {030365} (\bibinfo {year}
  {2025})}\BibitemShut {NoStop}%
\bibitem [{\citenamefont {Bolech}\ and\ \citenamefont
  {Demler}(2007)}]{bolech-07}%
  \BibitemOpen
  \bibfield  {author} {\bibinfo {author} {\bibfnamefont {C.~J.}\ \bibnamefont
  {Bolech}}\ and\ \bibinfo {author} {\bibfnamefont {E.}~\bibnamefont
  {Demler}},\ }\bibfield  {title} {\bibinfo {title} {Observing {M}ajorana bound
  states in p-wave superconductors using noise measurements in tunneling
  experiments},\ }\href {https://doi.org/10.1103/PhysRevLett.98.237002}
  {\bibfield  {journal} {\bibinfo  {journal} {Phys.\ Rev.\ Lett.}\ }\textbf
  {\bibinfo {volume} {98}},\ \bibinfo {pages} {237002} (\bibinfo {year}
  {2007})}\BibitemShut {NoStop}%
\bibitem [{\citenamefont {Law}\ \emph {et~al.}(2009)\citenamefont {Law},
  \citenamefont {Lee},\ and\ \citenamefont {Ng}}]{law-09}%
  \BibitemOpen
  \bibfield  {author} {\bibinfo {author} {\bibfnamefont {K.~T.}\ \bibnamefont
  {Law}}, \bibinfo {author} {\bibfnamefont {P.~A.}\ \bibnamefont {Lee}},\ and\
  \bibinfo {author} {\bibfnamefont {T.~K.}\ \bibnamefont {Ng}},\ }\bibfield
  {title} {\bibinfo {title} {{M}ajorana fermion induced resonant {A}ndreev
  reflection},\ }\href {https://doi.org/10.1103/PhysRevLett.103.237001}
  {\bibfield  {journal} {\bibinfo  {journal} {Phys.\ Rev.\ Lett.}\ }\textbf
  {\bibinfo {volume} {103}},\ \bibinfo {pages} {237001} (\bibinfo {year}
  {2009})}\BibitemShut {NoStop}%
\bibitem [{\citenamefont {Flensberg}(2010)}]{flensberg-10}%
  \BibitemOpen
  \bibfield  {author} {\bibinfo {author} {\bibfnamefont {K.}~\bibnamefont
  {Flensberg}},\ }\bibfield  {title} {\bibinfo {title} {Tunneling
  characteristics of a chain of {M}ajorana bound states},\ }\href
  {https://doi.org/10.1103/PhysRevB.82.180516} {\bibfield  {journal} {\bibinfo
  {journal} {Phys.\ Rev.\ B}\ }\textbf {\bibinfo {volume} {82}},\ \bibinfo
  {pages} {180516(R)} (\bibinfo {year} {2010})}\BibitemShut {NoStop}%
\bibitem [{\citenamefont {Vijay}\ and\ \citenamefont {Fu}(2016)}]{vijay-16}%
  \BibitemOpen
  \bibfield  {author} {\bibinfo {author} {\bibfnamefont {S.}~\bibnamefont
  {Vijay}}\ and\ \bibinfo {author} {\bibfnamefont {L.}~\bibnamefont {Fu}},\
  }\bibfield  {title} {\bibinfo {title} {Teleportation-based quantum
  information processing with {M}ajorana zero modes},\ }\href
  {https://doi.org/10.1103/PhysRevB.94.235446} {\bibfield  {journal} {\bibinfo
  {journal} {Phys. Rev. B}\ }\textbf {\bibinfo {volume} {94}},\ \bibinfo
  {pages} {235446} (\bibinfo {year} {2016})}\BibitemShut {NoStop}%
\bibitem [{\citenamefont {Karzig}\ \emph {et~al.}(2017)\citenamefont {Karzig},
  \citenamefont {Knapp}, \citenamefont {Lutchyn}, \citenamefont {Bonderson},
  \citenamefont {Hastings}, \citenamefont {Nayak}, \citenamefont {Alicea},
  \citenamefont {Flensberg}, \citenamefont {Plugge}, \citenamefont {Oreg},
  \citenamefont {Marcus},\ and\ \citenamefont {Freedman}}]{karzig-17}%
  \BibitemOpen
  \bibfield  {author} {\bibinfo {author} {\bibfnamefont {T.}~\bibnamefont
  {Karzig}}, \bibinfo {author} {\bibfnamefont {C.}~\bibnamefont {Knapp}},
  \bibinfo {author} {\bibfnamefont {R.~M.}\ \bibnamefont {Lutchyn}}, \bibinfo
  {author} {\bibfnamefont {P.}~\bibnamefont {Bonderson}}, \bibinfo {author}
  {\bibfnamefont {M.~B.}\ \bibnamefont {Hastings}}, \bibinfo {author}
  {\bibfnamefont {C.}~\bibnamefont {Nayak}}, \bibinfo {author} {\bibfnamefont
  {J.}~\bibnamefont {Alicea}}, \bibinfo {author} {\bibfnamefont
  {K.}~\bibnamefont {Flensberg}}, \bibinfo {author} {\bibfnamefont
  {S.}~\bibnamefont {Plugge}}, \bibinfo {author} {\bibfnamefont
  {Y.}~\bibnamefont {Oreg}}, \bibinfo {author} {\bibfnamefont {C.~M.}\
  \bibnamefont {Marcus}},\ and\ \bibinfo {author} {\bibfnamefont {M.~H.}\
  \bibnamefont {Freedman}},\ }\bibfield  {title} {\bibinfo {title} {Scalable
  designs for quasiparticle-poisoning-protected topological quantum computation
  with {M}ajorana zero modes},\ }\href
  {https://doi.org/10.1103/PhysRevB.95.235305} {\bibfield  {journal} {\bibinfo
  {journal} {Phys. Rev. B}\ }\textbf {\bibinfo {volume} {95}},\ \bibinfo
  {pages} {235305} (\bibinfo {year} {2017})}\BibitemShut {NoStop}%
\bibitem [{\citenamefont {Plugge}\ \emph {et~al.}(2017)\citenamefont {Plugge},
  \citenamefont {Rasmussen}, \citenamefont {Egger},\ and\ \citenamefont
  {Flensberg}}]{plugge-17}%
  \BibitemOpen
  \bibfield  {author} {\bibinfo {author} {\bibfnamefont {S.}~\bibnamefont
  {Plugge}}, \bibinfo {author} {\bibfnamefont {A.}~\bibnamefont {Rasmussen}},
  \bibinfo {author} {\bibfnamefont {R.}~\bibnamefont {Egger}},\ and\ \bibinfo
  {author} {\bibfnamefont {K.}~\bibnamefont {Flensberg}},\ }\bibfield  {title}
  {\bibinfo {title} {{M}ajorana box qubits},\ }\href
  {https://doi.org/10.1088/1367-2630/aa54e1} {\bibfield  {journal} {\bibinfo
  {journal} {New J. Phys.}\ }\textbf {\bibinfo {volume} {19}},\ \bibinfo
  {pages} {012001} (\bibinfo {year} {2017})}\BibitemShut {NoStop}%
\bibitem [{\citenamefont {Goldstein}\ and\ \citenamefont
  {Chamon}(2011)}]{goldstein-11}%
  \BibitemOpen
  \bibfield  {author} {\bibinfo {author} {\bibfnamefont {G.}~\bibnamefont
  {Goldstein}}\ and\ \bibinfo {author} {\bibfnamefont {C.}~\bibnamefont
  {Chamon}},\ }\bibfield  {title} {\bibinfo {title} {Decay rates for
  topological memories encoded with {M}ajorana fermions},\ }\href
  {https://doi.org/10.1103/physrevb.84.205109} {\bibfield  {journal} {\bibinfo
  {journal} {Phys. Rev. B}\ }\textbf {\bibinfo {volume} {84}},\ \bibinfo
  {pages} {205109} (\bibinfo {year} {2011})}\BibitemShut {NoStop}%
\bibitem [{\citenamefont {Budich}\ \emph {et~al.}(2012)\citenamefont {Budich},
  \citenamefont {Walter},\ and\ \citenamefont {Trauzettel}}]{budich-12}%
  \BibitemOpen
  \bibfield  {author} {\bibinfo {author} {\bibfnamefont {J.~C.}\ \bibnamefont
  {Budich}}, \bibinfo {author} {\bibfnamefont {S.}~\bibnamefont {Walter}},\
  and\ \bibinfo {author} {\bibfnamefont {B.}~\bibnamefont {Trauzettel}},\
  }\bibfield  {title} {\bibinfo {title} {Failure of protection of {M}ajorana
  based qubits against decoherence},\ }\href
  {https://doi.org/10.1103/physrevb.85.121405} {\bibfield  {journal} {\bibinfo
  {journal} {Phys. Rev. B}\ }\textbf {\bibinfo {volume} {85}},\ \bibinfo
  {pages} {121405(R)} (\bibinfo {year} {2012})}\BibitemShut {NoStop}%
\bibitem [{\citenamefont {Rainis}\ and\ \citenamefont
  {Loss}(2012)}]{rainis-12}%
  \BibitemOpen
  \bibfield  {author} {\bibinfo {author} {\bibfnamefont {D.}~\bibnamefont
  {Rainis}}\ and\ \bibinfo {author} {\bibfnamefont {D.}~\bibnamefont {Loss}},\
  }\bibfield  {title} {\bibinfo {title} {{M}ajorana qubit decoherence by
  quasiparticle poisoning},\ }\href
  {https://doi.org/10.1103/PhysRevB.85.174533} {\bibfield  {journal} {\bibinfo
  {journal} {Phys. Rev. B}\ }\textbf {\bibinfo {volume} {85}},\ \bibinfo
  {pages} {174533} (\bibinfo {year} {2012})}\BibitemShut {NoStop}%
\bibitem [{\citenamefont {Albrecht}\ \emph {et~al.}(2017)\citenamefont
  {Albrecht}, \citenamefont {Hansen}, \citenamefont {Higginbotham},
  \citenamefont {Kuemmeth}, \citenamefont {Jespersen}, \citenamefont {Nygard},
  \citenamefont {Krogstrup}, \citenamefont {Danon}, \citenamefont {Flensberg},\
  and\ \citenamefont {Marcus}}]{albrecht-17}%
  \BibitemOpen
  \bibfield  {author} {\bibinfo {author} {\bibfnamefont {S.~M.}\ \bibnamefont
  {Albrecht}}, \bibinfo {author} {\bibfnamefont {E.~B.}\ \bibnamefont
  {Hansen}}, \bibinfo {author} {\bibfnamefont {A.~P.}\ \bibnamefont
  {Higginbotham}}, \bibinfo {author} {\bibfnamefont {F.}~\bibnamefont
  {Kuemmeth}}, \bibinfo {author} {\bibfnamefont {T.~S.}\ \bibnamefont
  {Jespersen}}, \bibinfo {author} {\bibfnamefont {J.}~\bibnamefont {Nygard}},
  \bibinfo {author} {\bibfnamefont {P.}~\bibnamefont {Krogstrup}}, \bibinfo
  {author} {\bibfnamefont {J.}~\bibnamefont {Danon}}, \bibinfo {author}
  {\bibfnamefont {K.}~\bibnamefont {Flensberg}},\ and\ \bibinfo {author}
  {\bibfnamefont {C.~M.}\ \bibnamefont {Marcus}},\ }\bibfield  {title}
  {\bibinfo {title} {Transport signatures of quasiparticle poisoning in a
  {M}ajorana island},\ }\href {https://doi.org/10.1103/PhysRevLett.118.137701}
  {\bibfield  {journal} {\bibinfo  {journal} {Phys. Rev. Lett.}\ }\textbf
  {\bibinfo {volume} {118}},\ \bibinfo {pages} {137701} (\bibinfo {year}
  {2017})}\BibitemShut {NoStop}%
\bibitem [{\citenamefont {Karzig}\ \emph {et~al.}(2021)\citenamefont {Karzig},
  \citenamefont {Cole},\ and\ \citenamefont {Pikulin}}]{karzig-21}%
  \BibitemOpen
  \bibfield  {author} {\bibinfo {author} {\bibfnamefont {T.}~\bibnamefont
  {Karzig}}, \bibinfo {author} {\bibfnamefont {W.~S.}\ \bibnamefont {Cole}},\
  and\ \bibinfo {author} {\bibfnamefont {D.~I.}\ \bibnamefont {Pikulin}},\
  }\bibfield  {title} {\bibinfo {title} {Quasiparticle poisoning of {M}ajorana
  qubits},\ }\href {https://doi.org/10.1103/PhysRevLett.126.057702} {\bibfield
  {journal} {\bibinfo  {journal} {Phys. Rev. Lett.}\ }\textbf {\bibinfo
  {volume} {126}},\ \bibinfo {pages} {057702} (\bibinfo {year}
  {2021})}\BibitemShut {NoStop}%
\bibitem [{\citenamefont {Bhattacharyya}\ and\ \citenamefont {van
  Heck}(2026)}]{bhattacharyya-26}%
  \BibitemOpen
  \bibfield  {author} {\bibinfo {author} {\bibfnamefont {S.}~\bibnamefont
  {Bhattacharyya}}\ and\ \bibinfo {author} {\bibfnamefont {B.}~\bibnamefont
  {van Heck}},\ }\href {https://doi.org/10.48550/arXiv.2608.18042} {\bibinfo
  {title} {Dynamics of {M}ajorana tetron qubits under quasiparticle poisoning}}
  (\bibinfo {year} {2026}),\ \Eprint {https://arxiv.org/abs/arXiv:2608.18042}
  {arXiv:2608.18042} \BibitemShut {NoStop}%
\bibitem [{\citenamefont {Diehl}\ \emph {et~al.}(2011)\citenamefont {Diehl},
  \citenamefont {Rico}, \citenamefont {Baranov},\ and\ \citenamefont
  {Zoller}}]{diehl-11}%
  \BibitemOpen
  \bibfield  {author} {\bibinfo {author} {\bibfnamefont {S.}~\bibnamefont
  {Diehl}}, \bibinfo {author} {\bibfnamefont {E.}~\bibnamefont {Rico}},
  \bibinfo {author} {\bibfnamefont {M.~A.}\ \bibnamefont {Baranov}},\ and\
  \bibinfo {author} {\bibfnamefont {P.}~\bibnamefont {Zoller}},\ }\bibfield
  {title} {\bibinfo {title} {Topology by dissipation in atomic quantum wires},\
  }\href {https://doi.org/10.1038/nphys2106} {\bibfield  {journal} {\bibinfo
  {journal} {Nat. Phys.}\ }\textbf {\bibinfo {volume} {7}},\ \bibinfo {pages}
  {971} (\bibinfo {year} {2011})}\BibitemShut {NoStop}%
\bibitem [{\citenamefont {Krauter}\ \emph {et~al.}(2011)\citenamefont
  {Krauter}, \citenamefont {Muschik}, \citenamefont {Jensen}, \citenamefont
  {Wasilewski}, \citenamefont {Petersen}, \citenamefont {Cirac},\ and\
  \citenamefont {Polzik}}]{krauter-11}%
  \BibitemOpen
  \bibfield  {author} {\bibinfo {author} {\bibfnamefont {H.}~\bibnamefont
  {Krauter}}, \bibinfo {author} {\bibfnamefont {C.~A.}\ \bibnamefont
  {Muschik}}, \bibinfo {author} {\bibfnamefont {K.}~\bibnamefont {Jensen}},
  \bibinfo {author} {\bibfnamefont {W.}~\bibnamefont {Wasilewski}}, \bibinfo
  {author} {\bibfnamefont {J.~M.}\ \bibnamefont {Petersen}}, \bibinfo {author}
  {\bibfnamefont {J.~I.}\ \bibnamefont {Cirac}},\ and\ \bibinfo {author}
  {\bibfnamefont {E.~S.}\ \bibnamefont {Polzik}},\ }\bibfield  {title}
  {\bibinfo {title} {Entanglement generated by dissipation and steady state
  entanglement of two macroscopic objects},\ }\href
  {https://doi.org/10.1103/PhysRevLett.107.080503} {\bibfield  {journal}
  {\bibinfo  {journal} {Phys. Rev. Lett.}\ }\textbf {\bibinfo {volume} {107}},\
  \bibinfo {pages} {080503} (\bibinfo {year} {2011})}\BibitemShut {NoStop}%
\bibitem [{\citenamefont {Bardyn}\ \emph {et~al.}(2012)\citenamefont {Bardyn},
  \citenamefont {Baranov}, \citenamefont {Rico}, \citenamefont {İmamoğlu},
  \citenamefont {Zoller},\ and\ \citenamefont {Diehl}}]{bardyn-12}%
  \BibitemOpen
  \bibfield  {author} {\bibinfo {author} {\bibfnamefont {C.-E.}\ \bibnamefont
  {Bardyn}}, \bibinfo {author} {\bibfnamefont {M.~A.}\ \bibnamefont {Baranov}},
  \bibinfo {author} {\bibfnamefont {E.}~\bibnamefont {Rico}}, \bibinfo {author}
  {\bibfnamefont {A.}~\bibnamefont {İmamoğlu}}, \bibinfo {author}
  {\bibfnamefont {P.}~\bibnamefont {Zoller}},\ and\ \bibinfo {author}
  {\bibfnamefont {S.}~\bibnamefont {Diehl}},\ }\bibfield  {title} {\bibinfo
  {title} {{M}ajorana modes in driven-dissipative atomic superfluids with a
  zero {C}hern number},\ }\href
  {https://doi.org/10.1103/PhysRevLett.109.130402} {\bibfield  {journal}
  {\bibinfo  {journal} {Phys. Rev. Lett.}\ }\textbf {\bibinfo {volume} {109}},\
  \bibinfo {pages} {130402} (\bibinfo {year} {2012})}\BibitemShut {NoStop}%
\bibitem [{\citenamefont {Otterbach}\ and\ \citenamefont
  {Lemeshko}(2014)}]{otterbach-14}%
  \BibitemOpen
  \bibfield  {author} {\bibinfo {author} {\bibfnamefont {J.}~\bibnamefont
  {Otterbach}}\ and\ \bibinfo {author} {\bibfnamefont {M.}~\bibnamefont
  {Lemeshko}},\ }\bibfield  {title} {\bibinfo {title} {Dissipative preparation
  of spatial order in {R}ydberg-dressed {B}ose-{E}instein condensates},\ }\href
  {https://doi.org/10.1103/PhysRevLett.113.070401} {\bibfield  {journal}
  {\bibinfo  {journal} {Phys. Rev. Lett.}\ }\textbf {\bibinfo {volume} {113}},\
  \bibinfo {pages} {070401} (\bibinfo {year} {2014})}\BibitemShut {NoStop}%
\bibitem [{\citenamefont {Gong}\ \emph {et~al.}(2017)\citenamefont {Gong},
  \citenamefont {Higashikawa},\ and\ \citenamefont {Ueda}}]{gong-17}%
  \BibitemOpen
  \bibfield  {author} {\bibinfo {author} {\bibfnamefont {Z.}~\bibnamefont
  {Gong}}, \bibinfo {author} {\bibfnamefont {S.}~\bibnamefont {Higashikawa}},\
  and\ \bibinfo {author} {\bibfnamefont {M.}~\bibnamefont {Ueda}},\ }\bibfield
  {title} {\bibinfo {title} {{Z}eno {H}all effect},\ }\href
  {https://doi.org/10.1103/PhysRevLett.118.200401} {\bibfield  {journal}
  {\bibinfo  {journal} {Phys. Rev. Lett.}\ }\textbf {\bibinfo {volume} {118}},\
  \bibinfo {pages} {200401} (\bibinfo {year} {2017})}\BibitemShut {NoStop}%
\bibitem [{\citenamefont {Goldstein}(2019)}]{goldstein-19}%
  \BibitemOpen
  \bibfield  {author} {\bibinfo {author} {\bibfnamefont {M.}~\bibnamefont
  {Goldstein}},\ }\bibfield  {title} {\bibinfo {title} {Dissipation-induced
  topological insulators: A no-go theorem and a recipe},\ }\href
  {https://doi.org/10.21468/scipostphys.7.5.067} {\bibfield  {journal}
  {\bibinfo  {journal} {SciPost Physics}\ }\textbf {\bibinfo {volume} {7}},\
  \bibinfo {pages} {1} (\bibinfo {year} {2019})}\BibitemShut {NoStop}%
\bibitem [{\citenamefont {Gogoi}\ \emph {et~al.}(2026)\citenamefont {Gogoi},
  \citenamefont {Nag},\ and\ \citenamefont {Ghosh}}]{gogoi-26}%
  \BibitemOpen
  \bibfield  {author} {\bibinfo {author} {\bibfnamefont {K.}~\bibnamefont
  {Gogoi}}, \bibinfo {author} {\bibfnamefont {T.}~\bibnamefont {Nag}},\ and\
  \bibinfo {author} {\bibfnamefont {A.~K.}\ \bibnamefont {Ghosh}},\ }\bibfield
  {title} {\bibinfo {title} {Dissipation induced {M}ajarona $0-$ and
  $\pi$-modes in a driven {R}ashba nanowire},\ }\href
  {https://doi.org/10.1103/fxs7-wdqp} {\bibfield  {journal} {\bibinfo
  {journal} {Phys. Rev. B}\ }\textbf {\bibinfo {volume} {113}},\ \bibinfo
  {pages} {054301} (\bibinfo {year} {2026})}\BibitemShut {NoStop}%
\bibitem [{\citenamefont {Stern}\ \emph {et~al.}(1990)\citenamefont {Stern},
  \citenamefont {Aharonov},\ and\ \citenamefont {Imry}}]{stern-90}%
  \BibitemOpen
  \bibfield  {author} {\bibinfo {author} {\bibfnamefont {A.}~\bibnamefont
  {Stern}}, \bibinfo {author} {\bibfnamefont {Y.}~\bibnamefont {Aharonov}},\
  and\ \bibinfo {author} {\bibfnamefont {Y.}~\bibnamefont {Imry}},\ }\bibfield
  {title} {\bibinfo {title} {Phase uncertainty and loss of interference: A
  general picture},\ }\href {https://doi.org/10.1103/PhysRevA.41.3436}
  {\bibfield  {journal} {\bibinfo  {journal} {Phys. Rev. A}\ }\textbf {\bibinfo
  {volume} {41}},\ \bibinfo {pages} {3436} (\bibinfo {year}
  {1990})}\BibitemShut {NoStop}%
\bibitem [{\citenamefont {Datta}(1995)}]{datta-95}%
  \BibitemOpen
  \bibfield  {author} {\bibinfo {author} {\bibfnamefont {S.}~\bibnamefont
  {Datta}},\ }\href@noop {} {}Electronic transport in mesoscopic systems\
  (\bibinfo  {publisher} {Cambridge University Press, New York},\ \bibinfo
  {year} {1995})\BibitemShut {NoStop}%
\bibitem [{\citenamefont {D'Abbruzzo}\ and\ \citenamefont
  {Rossini}(2021)}]{dabbruzzo-21a}%
  \BibitemOpen
  \bibfield  {author} {\bibinfo {author} {\bibfnamefont {A.}~\bibnamefont
  {D'Abbruzzo}}\ and\ \bibinfo {author} {\bibfnamefont {D.}~\bibnamefont
  {Rossini}},\ }\bibfield  {title} {\bibinfo {title} {Self-consistent micro-
  scopic derivation of {M}arkovian master equations for open quadratic quantum
  systems},\ }\href {https://doi.org/10.1103/PhysRevA.103.052209} {\bibfield
  {journal} {\bibinfo  {journal} {Phys. Rev. A}\ }\textbf {\bibinfo {volume}
  {103}},\ \bibinfo {pages} {052209} (\bibinfo {year} {2021})}\BibitemShut
  {NoStop}%
\bibitem [{\citenamefont {Shustin}\ \emph {et~al.}(2026)\citenamefont
  {Shustin}, \citenamefont {Aksenov},\ and\ \citenamefont
  {Burmistrov}}]{shustin-26}%
  \BibitemOpen
  \bibfield  {author} {\bibinfo {author} {\bibfnamefont {M.~S.}\ \bibnamefont
  {Shustin}}, \bibinfo {author} {\bibfnamefont {S.~V.}\ \bibnamefont
  {Aksenov}},\ and\ \bibinfo {author} {\bibfnamefont {I.~S.}\ \bibnamefont
  {Burmistrov}},\ }\bibfield  {title} {\bibinfo {title} {Dissipation-induced
  steady states in topological superconductors: Mechanisms and design
  principles},\ }\href {https://doi.org/10.21468/SciPostPhys.20.5.138}
  {\bibfield  {journal} {\bibinfo  {journal} {SciPost Phys.}\ }\textbf
  {\bibinfo {volume} {20}},\ \bibinfo {pages} {138} (\bibinfo {year}
  {2026})}\BibitemShut {NoStop}%
\bibitem [{\citenamefont {Budich}\ and\ \citenamefont
  {Ardonne}(2013)}]{Budich2013}%
  \BibitemOpen
  \bibfield  {author} {\bibinfo {author} {\bibfnamefont {J.~C.}\ \bibnamefont
  {Budich}}\ and\ \bibinfo {author} {\bibfnamefont {E.}~\bibnamefont
  {Ardonne}},\ }\bibfield  {title} {\bibinfo {title} {Equivalent topological
  invariants for one-dimensional {M}ajorana wires in symmetry class},\ }\href
  {https://doi.org/10.1103/physrevb.88.075419} {\bibfield  {journal} {\bibinfo
  {journal} {Phys. Rev. B}\ }\textbf {\bibinfo {volume} {88}},\ \bibinfo
  {pages} {075419} (\bibinfo {year} {2013})}\BibitemShut {NoStop}%
\bibitem [{\citenamefont {Sato}\ and\ \citenamefont {Ando}(2017)}]{Sato2017}%
  \BibitemOpen
  \bibfield  {author} {\bibinfo {author} {\bibfnamefont {M.}~\bibnamefont
  {Sato}}\ and\ \bibinfo {author} {\bibfnamefont {Y.}~\bibnamefont {Ando}},\
  }\bibfield  {title} {\bibinfo {title} {Topological superconductors: a
  review},\ }\href {https://doi.org/10.1088/1361-6633/aa6ac7} {\bibfield
  {journal} {\bibinfo  {journal} {Reports on Progress in Physics}\ }\textbf
  {\bibinfo {volume} {80}},\ \bibinfo {pages} {076501} (\bibinfo {year}
  {2017})}\BibitemShut {NoStop}%
\bibitem [{\citenamefont {Valkov}\ \emph {et~al.}(2022)\citenamefont {Valkov},
  \citenamefont {Shustin}, \citenamefont {Aksenov}, \citenamefont {Zlotnikov},
  \citenamefont {Fedoseev}, \citenamefont {Mitskan},\ and\ \citenamefont
  {Kagan}}]{valkov-22}%
  \BibitemOpen
  \bibfield  {author} {\bibinfo {author} {\bibfnamefont {V.}~\bibnamefont
  {Valkov}}, \bibinfo {author} {\bibfnamefont {M.}~\bibnamefont {Shustin}},
  \bibinfo {author} {\bibfnamefont {S.}~\bibnamefont {Aksenov}}, \bibinfo
  {author} {\bibfnamefont {A.}~\bibnamefont {Zlotnikov}}, \bibinfo {author}
  {\bibfnamefont {A.}~\bibnamefont {Fedoseev}}, \bibinfo {author}
  {\bibfnamefont {V.}~\bibnamefont {Mitskan}},\ and\ \bibinfo {author}
  {\bibfnamefont {M.}~\bibnamefont {Kagan}},\ }\bibfield  {title} {\bibinfo
  {title} {Topological superconductivity and {M}ajorana states in
  low-dimensional systems},\ }\href
  {https://doi.org/10.3367/UFNe.2021.03.038950} {\bibfield  {journal} {\bibinfo
   {journal} {Phys. Usp.}\ }\textbf {\bibinfo {volume} {65}},\ \bibinfo {pages}
  {2} (\bibinfo {year} {2022})}\BibitemShut {NoStop}%
\bibitem [{\citenamefont {Aksenov}\ \emph {et~al.}(2026)\citenamefont
  {Aksenov}, \citenamefont {Shustin},\ and\ \citenamefont
  {Burmistrov}}]{aksenov-26}%
  \BibitemOpen
  \bibfield  {author} {\bibinfo {author} {\bibfnamefont {S.~V.}\ \bibnamefont
  {Aksenov}}, \bibinfo {author} {\bibfnamefont {M.~S.}\ \bibnamefont
  {Shustin}},\ and\ \bibinfo {author} {\bibfnamefont {I.~S.}\ \bibnamefont
  {Burmistrov}},\ }\bibfield  {title} {\bibinfo {title} {Controlling quantum
  transport in a superconducting device via dissipative baths},\ }\href
  {https://doi.org/10.1103/2361-9rxg} {\bibfield  {journal} {\bibinfo
  {journal} {Phys. Rev. B}\ }\textbf {\bibinfo {volume} {113}},\ \bibinfo
  {pages} {075428} (\bibinfo {year} {2026})}\BibitemShut {NoStop}%
\bibitem [{\citenamefont {Nilsson}\ \emph {et~al.}(2008)\citenamefont
  {Nilsson}, \citenamefont {Akhmerov},\ and\ \citenamefont
  {Beenakker}}]{nilsson-08}%
  \BibitemOpen
  \bibfield  {author} {\bibinfo {author} {\bibfnamefont {J.}~\bibnamefont
  {Nilsson}}, \bibinfo {author} {\bibfnamefont {A.~R.}\ \bibnamefont
  {Akhmerov}},\ and\ \bibinfo {author} {\bibfnamefont {C.~W.~J.}\ \bibnamefont
  {Beenakker}},\ }\bibfield  {title} {\bibinfo {title} {Splitting of a {C}ooper
  pair by a pair of {M}ajorana bound states},\ }\href
  {https://doi.org/10.1103/PhysRevLett.101.120403} {\bibfield  {journal}
  {\bibinfo  {journal} {Phys.\ Rev.\ Lett.}\ }\textbf {\bibinfo {volume}
  {101}},\ \bibinfo {pages} {120403} (\bibinfo {year} {2008})}\BibitemShut
  {NoStop}%
\bibitem [{\citenamefont {Tikhonov}\ and\ \citenamefont
  {Khrapai}(2026)}]{tikhonov-26}%
  \BibitemOpen
  \bibfield  {author} {\bibinfo {author} {\bibfnamefont {E.~S.}\ \bibnamefont
  {Tikhonov}}\ and\ \bibinfo {author} {\bibfnamefont {V.~S.}\ \bibnamefont
  {Khrapai}},\ }\bibfield  {title} {\bibinfo {title} {Conductance measurements
  cannot distinguish crossed {A}ndreev reflection from elastic co-tunnelling in
  normal–superconductor–normal junctions},\ }\href
  {https://doi.org/10.1038/s41567-026-03377-2} {\bibfield  {journal} {\bibinfo
  {journal} {Nat. Phys.}\ }\textbf {\bibinfo {volume} {22}},\ \bibinfo {pages}
  {1198} (\bibinfo {year} {2026})}\BibitemShut {NoStop}%
\bibitem [{\citenamefont {Vidal}\ \emph {et~al.}(2003)\citenamefont {Vidal},
  \citenamefont {Latorre}, \citenamefont {Rico},\ and\ \citenamefont
  {Kitaev}}]{vidal-03}%
  \BibitemOpen
  \bibfield  {author} {\bibinfo {author} {\bibfnamefont {G.}~\bibnamefont
  {Vidal}}, \bibinfo {author} {\bibfnamefont {J.~I.}\ \bibnamefont {Latorre}},
  \bibinfo {author} {\bibfnamefont {E.}~\bibnamefont {Rico}},\ and\ \bibinfo
  {author} {\bibfnamefont {A.}~\bibnamefont {Kitaev}},\ }\bibfield  {title}
  {\bibinfo {title} {Entanglement in quantum critical phenomena},\ }\href
  {https://doi.org/10.1103/physrevlett.90.227902} {\bibfield  {journal}
  {\bibinfo  {journal} {Phys. Rev. Lett.}\ }\textbf {\bibinfo {volume} {90}},\
  \bibinfo {pages} {075102} (\bibinfo {year} {2003})}\BibitemShut {NoStop}%
\bibitem [{\citenamefont {Turner}\ \emph {et~al.}(2011)\citenamefont {Turner},
  \citenamefont {Pollmann},\ and\ \citenamefont {Berg}}]{turner-11}%
  \BibitemOpen
  \bibfield  {author} {\bibinfo {author} {\bibfnamefont {A.~M.}\ \bibnamefont
  {Turner}}, \bibinfo {author} {\bibfnamefont {F.}~\bibnamefont {Pollmann}},\
  and\ \bibinfo {author} {\bibfnamefont {E.}~\bibnamefont {Berg}},\ }\bibfield
  {title} {\bibinfo {title} {Topological phases of one-dimensional fermions: An
  entanglement point of view},\ }\href
  {https://doi.org/10.1103/physrevb.83.075102} {\bibfield  {journal} {\bibinfo
  {journal} {Phys. Rev. B}\ }\textbf {\bibinfo {volume} {83}},\ \bibinfo
  {pages} {075102} (\bibinfo {year} {2011})}\BibitemShut {NoStop}%
\bibitem [{\citenamefont {Wu}\ and\ \citenamefont {Cao}(2012)}]{wu-12}%
  \BibitemOpen
  \bibfield  {author} {\bibinfo {author} {\bibfnamefont {B.~H.}\ \bibnamefont
  {Wu}}\ and\ \bibinfo {author} {\bibfnamefont {J.~C.}\ \bibnamefont {Cao}},\
  }\bibfield  {title} {\bibinfo {title} {Tunneling transport through
  superconducting wires with {M}ajorana bound states},\ }\href
  {https://doi.org/10.1103/PhysRevB.85.085415} {\bibfield  {journal} {\bibinfo
  {journal} {Phys.\ Rev.\ B}\ }\textbf {\bibinfo {volume} {85}},\ \bibinfo
  {pages} {085415} (\bibinfo {year} {2012})}\BibitemShut {NoStop}%
\bibitem [{\citenamefont {Tsintzis}\ \emph
  {et~al.}(2022{\natexlab{b}})\citenamefont {Tsintzis}, \citenamefont {Souto},\
  and\ \citenamefont {Leijnse}}]{tsintzis-22}%
  \BibitemOpen
  \bibfield  {author} {\bibinfo {author} {\bibfnamefont {A.}~\bibnamefont
  {Tsintzis}}, \bibinfo {author} {\bibfnamefont {R.~S.}\ \bibnamefont
  {Souto}},\ and\ \bibinfo {author} {\bibfnamefont {M.}~\bibnamefont
  {Leijnse}},\ }\bibfield  {title} {\bibinfo {title} {Creating and detecting
  poor man's {M}ajorana bound states in interacting quantum dots},\ }\href
  {https://doi.org/10.1103/PhysRevB.106.L201404} {\bibfield  {journal}
  {\bibinfo  {journal} {Phys. Rev. B}\ }\textbf {\bibinfo {volume} {106}},\
  \bibinfo {pages} {L201404} (\bibinfo {year}
  {2022}{\natexlab{b}})}\BibitemShut {NoStop}%
\bibitem [{\citenamefont {Pothier}\ \emph {et~al.}(1997)\citenamefont
  {Pothier}, \citenamefont {Guéron}, \citenamefont {Birge}, \citenamefont
  {Esteve},\ and\ \citenamefont {Devoret}}]{pothier-97}%
  \BibitemOpen
  \bibfield  {author} {\bibinfo {author} {\bibfnamefont {H.}~\bibnamefont
  {Pothier}}, \bibinfo {author} {\bibfnamefont {S.}~\bibnamefont {Guéron}},
  \bibinfo {author} {\bibfnamefont {N.~O.}\ \bibnamefont {Birge}}, \bibinfo
  {author} {\bibfnamefont {D.}~\bibnamefont {Esteve}},\ and\ \bibinfo {author}
  {\bibfnamefont {M.~H.}\ \bibnamefont {Devoret}},\ }\bibfield  {title}
  {\bibinfo {title} {Energy distribution function of quasiparticles in
  mesoscopic wires},\ }\href {https://doi.org/10.1103/PhysRevLett.79.3490}
  {\bibfield  {journal} {\bibinfo  {journal} {Phys. Rev. Lett.}\ }\textbf
  {\bibinfo {volume} {79}},\ \bibinfo {pages} {3490} (\bibinfo {year}
  {1997})}\BibitemShut {NoStop}%
\bibitem [{\citenamefont {Giazotto}\ \emph {et~al.}(2006)\citenamefont
  {Giazotto}, \citenamefont {Heikkilä}, \citenamefont {Luukanen},
  \citenamefont {Savin},\ and\ \citenamefont {Pekola}}]{giazotto-06}%
  \BibitemOpen
  \bibfield  {author} {\bibinfo {author} {\bibfnamefont {F.}~\bibnamefont
  {Giazotto}}, \bibinfo {author} {\bibfnamefont {T.~T.}\ \bibnamefont
  {Heikkilä}}, \bibinfo {author} {\bibfnamefont {A.}~\bibnamefont {Luukanen}},
  \bibinfo {author} {\bibfnamefont {A.~M.}\ \bibnamefont {Savin}},\ and\
  \bibinfo {author} {\bibfnamefont {J.~P.}\ \bibnamefont {Pekola}},\ }\bibfield
   {title} {\bibinfo {title} {Opportunities for mesoscopics in thermometry and
  refrigeration: Physics and applications},\ }\href
  {https://doi.org/10.1103/RevModPhys.78.217} {\bibfield  {journal} {\bibinfo
  {journal} {Rev. Mod. Phys.}\ }\textbf {\bibinfo {volume} {78}},\ \bibinfo
  {pages} {217} (\bibinfo {year} {2006})}\BibitemShut {NoStop}%
\bibitem [{\citenamefont {Breuer}\ and\ \citenamefont
  {Petruccione}(1996)}]{breuer-02}%
  \BibitemOpen
  \bibfield  {author} {\bibinfo {author} {\bibfnamefont {H.-P.}\ \bibnamefont
  {Breuer}}\ and\ \bibinfo {author} {\bibfnamefont {F.}~\bibnamefont
  {Petruccione}},\ }\href@noop {} {}The theory of open quantum systems\
  (\bibinfo  {publisher} {Oxford University Press, Oxford},\ \bibinfo {year}
  {1996})\BibitemShut {NoStop}%
\bibitem [{\citenamefont {Dorda}\ \emph {et~al.}(2017)\citenamefont {Dorda},
  \citenamefont {Sorantin}, \citenamefont {von~der Linden},\ and\ \citenamefont
  {Arrigoni}}]{dorda-17}%
  \BibitemOpen
  \bibfield  {author} {\bibinfo {author} {\bibfnamefont {A.}~\bibnamefont
  {Dorda}}, \bibinfo {author} {\bibfnamefont {M.}~\bibnamefont {Sorantin}},
  \bibinfo {author} {\bibfnamefont {W.}~\bibnamefont {von~der Linden}},\ and\
  \bibinfo {author} {\bibfnamefont {E.}~\bibnamefont {Arrigoni}},\ }\bibfield
  {title} {\bibinfo {title} {Optimized auxiliary representation of
  non-{M}arkovian impurity problems by a {L}indblad equation},\ }\href
  {https://doi.org/10.1088/1367-2630/aa6ccc} {\bibfield  {journal} {\bibinfo
  {journal} {New J. Phys.}\ }\textbf {\bibinfo {volume} {19}},\ \bibinfo
  {pages} {063005} (\bibinfo {year} {2017})}\BibitemShut {NoStop}%
\bibitem [{\citenamefont {Jin}\ \emph {et~al.}(2020)\citenamefont {Jin},
  \citenamefont {Filippone},\ and\ \citenamefont {Giamarchi}}]{jin-20}%
  \BibitemOpen
  \bibfield  {author} {\bibinfo {author} {\bibfnamefont {T.}~\bibnamefont
  {Jin}}, \bibinfo {author} {\bibfnamefont {M.}~\bibnamefont {Filippone}},\
  and\ \bibinfo {author} {\bibfnamefont {T.}~\bibnamefont {Giamarchi}},\
  }\bibfield  {title} {\bibinfo {title} {Generic transport formula for a system
  driven by {M}arkovian reservoirs},\ }\href
  {https://doi.org/10.1103/PhysRevB.102.205131} {\bibfield  {journal} {\bibinfo
   {journal} {Phys. Rev. B}\ }\textbf {\bibinfo {volume} {102}},\ \bibinfo
  {pages} {205131} (\bibinfo {year} {2020})}\BibitemShut {NoStop}%
\bibitem [{\citenamefont {Thompson}\ and\ \citenamefont
  {Kamenev}(2023)}]{thompson-23}%
  \BibitemOpen
  \bibfield  {author} {\bibinfo {author} {\bibfnamefont {F.}~\bibnamefont
  {Thompson}}\ and\ \bibinfo {author} {\bibfnamefont {A.}~\bibnamefont
  {Kamenev}},\ }\bibfield  {title} {\bibinfo {title} {Field theory of many-body
  {L}indbladian dynamics},\ }\href {https://doi.org/10.1016/j.aop.2023.169385}
  {\bibfield  {journal} {\bibinfo  {journal} {Ann. Phys. (N.Y.)}\ }\textbf
  {\bibinfo {volume} {455}},\ \bibinfo {pages} {169385} (\bibinfo {year}
  {2023})}\BibitemShut {NoStop}%
\bibitem [{\citenamefont {Surace}\ and\ \citenamefont
  {Tagliacozzo}(2022)}]{surace-22}%
  \BibitemOpen
  \bibfield  {author} {\bibinfo {author} {\bibfnamefont {J.}~\bibnamefont
  {Surace}}\ and\ \bibinfo {author} {\bibfnamefont {L.}~\bibnamefont
  {Tagliacozzo}},\ }\bibfield  {title} {\bibinfo {title} {Fermionic {G}aussian
  states: {An} introduction to numerical approaches},\ }\href
  {https://doi.org/10.21468/SciPostPhysLectNotes.54} {\bibfield  {journal}
  {\bibinfo  {journal} {SciPost Physics Lecture Notes}\ }\textbf {\bibinfo
  {volume} {54}},\ \bibinfo {pages} {1} (\bibinfo {year} {2022})}\BibitemShut
  {NoStop}%
\bibitem [{\citenamefont {Grabsch}\ \emph {et~al.}(2019)\citenamefont
  {Grabsch}, \citenamefont {Cheipesh},\ and\ \citenamefont
  {Beenakker}}]{grabsch-19}%
  \BibitemOpen
  \bibfield  {author} {\bibinfo {author} {\bibfnamefont {A.}~\bibnamefont
  {Grabsch}}, \bibinfo {author} {\bibfnamefont {Y.}~\bibnamefont {Cheipesh}},\
  and\ \bibinfo {author} {\bibfnamefont {C.~W.~J.}\ \bibnamefont {Beenakker}},\
  }\bibfield  {title} {\bibinfo {title} {Pfaffian formula for fermion parity
  fluctuations in a superconductor and application to {M}ajorana fusion
  detection},\ }\href {https://doi.org/10.1002/andp.201900129} {\bibfield
  {journal} {\bibinfo  {journal} {Ann. Phys. (Berlin)}\ }\textbf {\bibinfo
  {volume} {531}},\ \bibinfo {pages} {1900129} (\bibinfo {year}
  {2019})}\BibitemShut {NoStop}%
\end{thebibliography}%

\vspace{0.2cm}
\onecolumngrid
    \begin{center} 
    {\bf \large End Matter}
\end{center}

\twocolumngrid

\setcounter{equation}{0}
\renewcommand{\theequation}{A\arabic{equation}}

\noindent\textsf{\color{blue} Appendix A. MBS in isolated Kitaev chain.} To diagonalize 
$\hat{H}_s$, it is convenient to switch to the 
Majorana representation: $\hat{w}_{2n-1}=\hat{c}_n+\hat{c}_n^\dagger$ and 
$\hat{w}_{2n}=i(\hat{c}_n^\dagger-\hat{c}_n)$. We then find 
$\hat{H}_s = i \underline{w}^T A \underline{w}$, where $A$ is a $2N{\times}2N$ 
real skew-symmetric matrix, $A=A^*=-A^T$. It can be brought to canonical form 
by an orthogonal rotation:
\begin{equation}
A = \mathcal{W} \bigoplus_{a=1}^{N} \begin{pmatrix}
0 & \varepsilon_a \\
-\varepsilon_a & 0 
\end{pmatrix} \mathcal{W}^T, \quad \mathcal{W} \in O(2N).   
\end{equation}
The orthogonal matrix 
$\mathcal{W} = [\,\tilde{\underline{\chi}}_{1},~\tilde{\underline{\chi}}_{2},\ldots,~\tilde{\underline{\chi}}_{2N-1},~\tilde{\underline{\chi}}_{2N}\,]$  
is composed of $2N$ orthonormal vectors, $\tilde{\underline{\chi}}_{a} \cdot \tilde{\underline{\chi}}_{b}=\delta_{ab}$. Since for the Hamiltonian $\hat{H}_s$ in the BDI class the matrix $A$ has a block structure, the orthonormal vectors live on two different sublattices: 
$\tilde{\chi}_{2a-1,2n}=\tilde{\chi}_{2a,2n-1}\equiv 0$.
The eigenvalues 
$\varepsilon_a = \tilde{\underline{\chi}}_{2a-1}^T A \tilde{\underline{\chi}}_{2a}$ have the meaning of Bogoliubov quasiparticle energies for $\hat{H}_s$. The corresponding creation operators can be written as $\hat{\alpha}_a^\dagger = (\tilde{\underline{\chi}}_{2a-1}- i \tilde{\underline{\chi}}_{2a})\cdot \underline{\hat{w}}/2$. In the topological phase with a single MBS, there is a zero-energy quasiparticle, $\varepsilon_1=0$. The corresponding orthonormal vectors $\tilde{\underline{\chi}}_{1,2}$ are localized on different sublattices. Therefore, in the sublattice representation, instead of $\tilde{\underline{\chi}}_{1,2}$, we can work with vectors 
$\underline{\chi}_{1,N}\in \mathbb{R}^N$: 
$\underline{\chi}_{1,n} \equiv \underline{\tilde{\chi}}_{1,2n-1}$, $\underline{\chi}_{N,n} \equiv \underline{\tilde{\chi}}_{2,2n}$. These are the vectors used in the main text. The notation is motivated by the fact that at the sweet spot we have $\hat{H}_s=i t\sum_{n=1}^{N-1} \hat{w}_{2n} \hat{w}_{2n+1}$, so that $\hat{w}_1$ and $\hat{w}_{2N}$ commute with $\hat{H}_s$, i.e., the MBS orthonormal vectors are 
$\tilde{\chi}_{1,n}=\delta_{n,1}$ ($\underline{\chi}_{1,n}=\delta_{n,1}$) and $\tilde{\chi}_{2,n}=\delta_{n,2N}$ ($\underline{\chi}_{N,n}=\delta_{n,N}$).

\setcounter{equation}{0}
\renewcommand{\theequation}{B\arabic{equation}}
\noindent\textsf{\color{blue} Appendix B. Meir-Wingreen formula for dissipative system.} Following Ref.~\cite{aksenov-26}, the various contributions to the differential 
conductance in Eq.~\eqref{G_conts} can be expressed as
\begin{gather}
    G_{j}^{LAR} = T_{eh}^{(j,j)}\left(V_j\right)+T_{eh}^{(j,j)}\left(-V_j\right), \quad G_{j}^{EC} = T_{ee}^{(j,\bar{j})}\left(V_j\right) , \notag \\
    G_{j}^{CAR} = T_{eh}^{(j,\bar{j})}\left(V_j\right), \, G_{j}^{B} = \underrightarrow{T}^{(j)}\left(V_j\right)+\underleftarrow{T}^{(j)}\left(V_j\right) ,
\label{Gj} 
\end{gather}
where the corresponding probabilities are given by $T_{e\alpha}^{(j,p)}=\Tr\{\Gamma_{je}G_{e\alpha}\Gamma_{p\alpha}G_{e\alpha}^{\dagger}\}$, with $\alpha=e,h$, and 
$\underrightarrow{T}^{(j)} =\Tr\{\Gamma_{je}\underrightarrow{b}\underrightarrow{b}^{\dagger}\}$, 
$\underleftarrow{T}^{(j)} =\Tr\{\Gamma_{je}\underleftarrow{b}\underleftarrow{b}^{\dagger}\}$,
or, explicitly,
\begin{equation}\label{Ts}
			\begin{split}
				T_{eh}^{(i,j)}  & {=} \Gamma_{i}\Gamma_{j}\left|\left[G_{eh}\right]_{i,j}\right|^2, ~
				T_{ee}^{(j,\bar{j})}  {=} \Gamma_{1}\Gamma_{N}\left|\left[G_{ee}\right]_{j,\bar{j}}\right|^2, \\
				\underrightarrow{T}^{(j)} &  {=} \Gamma_{j}\left[\underrightarrow{b}\underrightarrow{b}^{\dagger}\right]_{j,j},\hspace{0.45cm}
				\underleftarrow{T}^{(j)}  {=} \Gamma_{j}\left[\underleftarrow{b}\underleftarrow{b}^{\dagger}\right]_{j,j}, \\
                \underrightarrow{b}  & {=} G_{ee}\,\underline{\mu}^{*}+G_{eh}\,\underline{\nu}^{*},\hspace{0.45cm} \underleftarrow{b}  {=} G_{ee}\,\underline{\nu}+G_{eh}\,\underline{\mu} .
			\end{split}
		\end{equation}
Here $G_{ee}$ and $G_{eh}$ are the blocks of the retarded Green's function of 
the Kitaev chain,
\begin{equation}
    G^{R}= (\omega - \hat{h}_{\rm eff})^{-1} = \begin{pmatrix}
        G_{ee} & G_{eh} \\
G_{he} & G_{hh}
    \end{pmatrix}, 
\end{equation}
in the Nambu space. The effective Hamiltonian matrix is 
$\hat{h}_{\rm eff}=\hat{h}_{s}-i(\hat{\Gamma}^{(1)}+\hat{\Gamma}^{(N)}+\hat{Q})/2$, where
\begin{equation}
 \hat{h}_s = \begin{pmatrix}
     \hat{\xi} & \hat{\Delta}^\dagger \\
    \hat{\Delta} & - \hat{\xi}^T
 \end{pmatrix}   ,
 \label{eq:appB:hs}
\end{equation}
with $\hat{\xi}_{nm}=-\mu\delta_{nm}+t \delta_{m,n+1}+t^*\delta_{n,m+1}$ and 
$\hat{\Delta}_{nm}=\Delta\left(\delta_{m,n+1} - \delta_{n,m+1}\right)$.
The broadening matrices for the normal leads, $\Gamma^{(j)}$, are block diagonal 
in Nambu space: $\Gamma^{(j)}=\diag\{\Gamma_{je},~\Gamma_{jh}\}$, with 
$\left[\Gamma_{j\alpha}\right]_{n,m}=\Gamma_{j}\delta_{n,j}\delta_{m,j}$. The 
dissipative broadening matrix is defined as \cite{thompson-23}
\begin{equation}
 \hat{Q} = \begin{pmatrix}
\underline{\mu}^* \underline{\mu}^T+ \underline{\nu}\underline{\nu}^\dagger & \underline{\mu}^* \underline{\nu}^T + \underline{\nu} \underline{\mu}^\dagger\\
\underline{\nu}^* \underline{\mu}^T + \underline{\mu} \underline{\nu}^\dagger & \underline{\mu} \underline{\mu}^\dagger+ \underline{\nu}^*\underline{\nu}^T
\end{pmatrix} .
\end{equation}

\setcounter{equation}{0}
\renewcommand{\theequation}{C\arabic{equation}}
\noindent\textsf{\color{blue} Appendix C. Projection to lowest-energy states.} To employ the Meir-Wingreen formula from Appendix B, we need to compute 
$G_{ee(h)}$. As in Appendix A, it is convenient to switch to the 
Majorana-operator basis. This transformation is achieved by a rotation with 
the block matrix
\begin{equation}\label{f2m}
\mathcal{U}=\begin{pmatrix}
1 & 1 \\
-i & i \\
\end{pmatrix} 
\end{equation}
in Nambu space. We then find $X=i\, \mathcal{U}h_{\rm eff}\,\mathcal{U}^{-1}/2=-2A+M_{r}$, where 
$M_{r}=\Re \big(\, \ell\ell^\dagger + \sum_{j=1,N} \mathtt{t}_{j}\mathtt{t}_{j}^\dagger\,\big)$
is a real symmetric matrix, $M_r=M_r^*=M_r^T$. Here, we introduce  
$\mathtt{t}_{j} = \underline{\mathtt{t}}_{j}^{r}+ i\underline{\mathtt{t}}_{j}^{i}$ and  
$\ell = \underline{\ell}^{r}+ i\underline{\ell}^{i}$. The vectors 
$\underline{\mathtt{t}}_{j}^{r(i)} \in \mathbb{R}^N$ have 
a few nonzero 
elements due to geometry of the setup:
\begin{equation}\label{tri}
\left[\underline{\mathtt{t}}_{j}^{r(i)}\right]_{n}  = \frac{\sqrt{2\pi\rho_{j}}}{2}\left(\delta_{n,j}t_{j}^{r(i)}\mp\delta_{n,N+j}t_{j}^{i(r)}\right).
\end{equation}
Note that $\underline{\mathtt{t}}_{j}^{r(i)}$ have a block structure in 
Majorana space. In particular, their first (last) $N$ elements correspond to 
the $A$ ($B$) Majorana sublattice, described by the operators 
$\hat{w}_{An}\equiv \hat{w}_{2n-1}$ and $\hat{w}_{Bn}\equiv \hat{w}_{2n}$. 
Thus, we can write
\begin{equation}\label{ttri}
			\underline{\mathtt{t}}_{j}^{r(i)} = \frac{\sqrt{2\pi\rho_{j}}}{2}
			\begin{pmatrix}
				\underline{t_{j}}^{r(i)} \\
				\mp \underline{t_{j}}^{i(r)}
			\end{pmatrix} ,
\end{equation}
where $\underline{t_{1}}^{r(i)}{=}\{t_1^{r(i)},0,\dots, 0\}$ and 
$\underline{t_{N}}^{r(i)}{=}\{0,\dots, 0,t_N^{r(i)}\}$. The vectors  
$\underline{\ell}^{r(i)} \in \mathbb{R}^{2N}$ are defined analogously,
\begin{equation}\label{lri}
			\underline{\ell}^{r(i)} = \frac{1}{2}
			\begin{pmatrix}
				(\underline{\mu}^{r(i)}+\underline{\nu}^{r(i)}) \\
				\mp (\underline{\mu}^{i(r)}-\underline{\nu}^{i(r)})
			\end{pmatrix}.
\end{equation}
Introducing the eigenvalues and eigenvectors of $X$,
\begin{equation}
\begin{split}
 X   R_a & = \beta_a  R_a, \quad R_a= \begin{pmatrix}
 \underline{r}_{aA} \\
			\underline{r}_{aB} 
            \end{pmatrix} ,
            \\
            X^T  L_a &  = \beta_a L_a, \quad L_a=\begin{pmatrix}
 \underline{l}_{aA} \\
			\underline{l}_{aB} 
            \end{pmatrix} ,
            \end{split}
\end{equation}
the spectral decomposition for the blocks $G_{ee(h)}$ 
becomes
\begin{equation}\label{GeeGeh_lr}
        	G_{ee(eh)}=\frac{1}{2}\sum\limits_{a=1}^{N}\frac{\left(\underline{r}_{aA}+i\underline{r}_{aB}\right) \left(\underline{l}_{aA}^T \mp i\underline{l}_{aB}^T\right)}{\omega+2i\beta_{a}} .
\end{equation}

To compute the conductance contributions from the lowest-energy states, we 
project $X$ onto the MBS subspace, introducing the $2\times 2$ matrix 
$\mathcal{X}$ with elements 
$\mathcal{X}_{ab}=\tilde{\underline{\chi}}_{a}^T X \tilde{\underline{\chi}}_{b}$. 
The real symmetric matrix $\mathcal{X}$ reads
\begin{equation}\label{X_eff} \mathcal{X}= 
\begin{pmatrix} 
g_1 + |\bm{b}_1|^2 & [{\bm{b}}_N \times {\bm{b}}_1] \\ 
[{\bm{b}}_N \times {\bm{b}}_1] & g_N + |\bm{b}_N|^2 
\end{pmatrix} .
\end{equation}
Its eigenvalues are $\beta_{1,N}$, as given in Eq.~\eqref{eq:beta:1N}. The 
corresponding right and left eigenvectors of $X$ can be approximated as
\begin{equation}
  r_{aA}=l_{aA} \simeq x_{1a} {\underline{\chi}_{1}}, \quad r_{aB}=l_{aB} \simeq x_{2a} {\underline{\chi}_{N}} ,    
\label{eq:C8}
\end{equation}
where 
\begin{align}
x_{12(21)} &= \pm \operatorname{sgn} [\textbf{b}_N\times \textbf{b}_1]\left [\frac{1}{2} - \frac{\sum_{j}s_j(g_j+|\bm{b}_j|^2)}{2(\beta_1-\beta_N)} 
\right ]^{1/2} , \notag \\
x_{11(22)} & = \left [\frac{1}{2}+ \frac{\sum_{j}s_j(g_j+|\bm{b}_j|^2)}{2(\beta_1-\beta_N)} 
\right ]^{1/2} . 
\end{align}
Using the above expressions, one can 
derive Eq.~\eqref{G_terms}.

\setcounter{equation}{0}
\renewcommand{\theequation}{D\arabic{equation}}
\noindent\textsf{\color{blue} Appendix D. Reduced density matrix.} Let us consider the approximate structure of the density matrix on the pair of low-energy many-body states corresponding to the empty and occupied MBS in the regime of the symmetric coupling to the leads, $g_{1,N}\equiv g > 0$, and zero bias voltage. Since in our system $\beta_{1,N} > 0$, the 
density matrix is Gaussian. As known \cite{surace-22}, it can be written as
\begin{equation}
\rho=\bigoplus_{a=1}^N
\begin{pmatrix}
(1+f_a)/2 & 0 \\
0 & (1-f_a)/2
\end{pmatrix},    
\end{equation}
where $f_a$ are the eigenvalues of the correlation matrix 
$F_{ij} = \Tr \rho [\,\hat{w}_i,\hat{w}_j]/2$, 
which satisfies the Lyapuno--Sylvester equation
$XF + FX^T - Y = 0$,
with the driving matrix $Y = -Y^* = -Y^T$  defined as
\begin{equation}
Y =  2i\,\Im \Big(\, \ell\ell^\dagger + \sum_{j=1,N} \mathtt{t}_{j}\mathtt{t}_{j}^\dagger\,\Big).
\end{equation}
Solving the Lyapunov--Sylvester equation yields $(R^T F L)_{ab} = (L^T Y L)_{ab}/(\beta_a+\beta_b)$.
 Restricting to the lowest-energy subspace and using Eq.~\eqref{eq:C8}, we find
\begin{equation}
    (R^T F L)_{ab} \to \frac{2(\bm{b}_1\cdot \bm{b}_N)}{\beta_1+\beta_N} \begin{pmatrix}
        0 & -i \\
        i & 0
    \end{pmatrix}.
\end{equation}
This result immediately implies that the relevant eigenvalue is 
$f_1=2(\bm{b}_1\cdot \bm{b}_N)/(\beta_1+\beta_N)$. For a symmetric setup, 
this gives Eq.~\eqref{rho} with $p_g\equiv f_1$.
We note that $f_1$ determines the parity of the occupation of the lowest-energy 
state, but not the fermion parity of the entire system. The latter is given by 
$\hat{P} = \exp(i\pi \sum_n c_n^\dagger c_n)$, so that 
$\langle \hat{P} \rangle = \operatorname{Pf}(F) = \prod_{a=1}^{N} f_a$ \cite{grabsch-19}.

\setcounter{equation}{0}
\renewcommand{\theequation}{E\arabic{equation}}
\noindent\textsf{\color{blue} Appendix E. AB setup.} To compute electric transport in the AB setup, one can use the same 
Meir-Wingreen-type expressions described in Appendix B after doubling the 
Hilbert space to include both arms of the interferometer. In particular, the 
Hamiltonian matrix $\hat{h}_s$ takes the form of Eq.~\eqref{eq:appB:hs} but with
\begin{equation}
 \hat{\xi} \to \begin{pmatrix}
  \hat{\xi} & 0 \\
  0 & \hat{\xi}_0
 \end{pmatrix}, \qquad    \hat{\Delta} \to \begin{pmatrix}
  \hat{\Delta} & 0 \\
  0 & 0
 \end{pmatrix} ,
\end{equation}
where $[\hat{\xi}_0]_{nm}=\xi_0 \delta_{nm}+t_0 \delta_{m,n+1}+t_0^*\delta_{n,m+1}$. 
Furthermore,
\begin{gather}
 \hat{\Gamma}^{(1)}+\hat{\Gamma}^{(N)} \to    \begin{pmatrix}
  \hat{\Gamma}^{(1)}+\hat{\Gamma}^{(N)} & \hat{\Gamma}_+^{(1)}+\hat{\Gamma}_+^{(N)} \\
  \hat{\Gamma}_-^{(1)}+\hat{\Gamma}_-^{(N)} & \hat{\Gamma}^{'(1)}+\hat{\Gamma}^{'(N)}
 \end{pmatrix} ,\notag \\
 \left[\Gamma_{\pm}^{(j)}\right]_{n,m} = \sqrt{\Gamma_{j}\Gamma_{j}'}e^{\pm i s_j \phi}\delta_{n,j}\delta_{m,j} .
\end{gather}
Here $\hat{\Gamma}^{'(j)}$ (with $\Gamma_{j}'$) is defined in the same way as 
$\hat{\Gamma}^{(j)}$ (with $\Gamma_{j}$), but with $t_j'$ substituted for 
$t_j$. Finally, we define
\begin{equation}
 \hat{Q} \to    \begin{pmatrix}
  \hat{Q} & 0 \\
  0 & 0
 \end{pmatrix} .
\end{equation}

\setcounter{equation}{0}
\renewcommand{\theequation}{F\arabic{equation}}
\noindent\textsf{\color{blue} Appendix F. Topology-by-dissipation model.} According to Theorem~1 of Ref.~\cite{shustin-26}, the number of zero 
eigenvalues of $X$ for quadratic Liouvillians is
$
N_0 = 2N_M - \operatorname{rk} \mathcal{B}.
$
The cross-Gram hybridization matrix $\mathcal{B}$ is constructed from the 
$2N_M$ Majorana wave functions 
$\underline{\tilde\chi}_{1},\ldots, \underline{\tilde\chi}_{2N_M} \in \ker A$ 
and the $2N_B$ dissipative fields 
$\underline{\ell}_{1}^{r,i}, \ldots, \underline{\ell}_{N_B}^{r,i}$. In the 
absence of unitary dynamics ($A=0$), we may take 
$\ker A = \operatorname{Span}\{\tilde{\underline{\chi}}_{1},\ldots,\tilde{\underline{\chi}}_{2N}\}$, 
with $\underline{\tilde\chi}_{a}$ defined in Appendix~A. Furthermore, we 
choose $N_B = N$, $\underline{\ell}_N \in \operatorname{Span}\{\tilde{\underline{\chi}}_{1},\,\tilde{\underline{\chi}}_{2}\}$, and $\underline{\ell}_{v} \in \operatorname{Span}\{\tilde{\underline{\chi}}_{3},\ldots,\tilde{\underline{\chi}}_{2N}\}$. Then we find $X = X_0 + \Re\big(\underline{\ell}_N\underline{\ell}_N^\dagger\big)$, where  $X_0 = \Re \sum_{v=1}^{N-1}\underline{\ell}_v\underline{\ell}_v^\dagger$.
By construction, $X_0$ has two zero modes ($N_0=2$) with eigenvectors in the 
subspace spanned by $\tilde{\underline{\chi}}_{1,2}$ and a spectral gap 
$\Delta\beta \sim \|\underline{\ell}_v\|^2$.  
Assuming $\|\ell_N\| \ll \|\ell_{v}\|$, we treat the 
the term $X-X_0$
perturbatively. Projecting this perturbation onto the subspace of 
$\underline{\tilde{\chi}}_{1,2}$ recovers the effective low-energy 
Hamiltonian \eqref{X_eff}, which governs the nonlocal transport properties 
discussed in the main text.

\end{document}